\documentclass[%
 reprint,
nofootinbib,
 amsmath,amssymb,
 aps,
prc,
]{revtex4-1}

\usepackage{graphicx}
\usepackage{dcolumn}
\usepackage{bm}
\usepackage{hyperref}
\usepackage{xcolor}

\newcommand{\rmd}{\mathrm{d}}

\newcommand{\rmi}{\mathrm{i}}

\newcommand*\PS[1]{\boldsymbol{p}_{#1}}			

\newcommand*\OPD[2]{ E_{\PS{{#1}}}\frac{\rmd N^{#2}}{\rmd^3 p_{{#1}}} }%
\newcommand*\TPD[3]{ E_{\PS{{#1}}} E_{\PS{{#2}}} \frac{\rmd N^{#3}}{\rmd^3 p_{{#1}}\rmd^3 p_{{#2}}} }%

\newcommand{\QS}{\boldsymbol{q}} 	
\newcommand{\QT}{\boldsymbol{q}_\perp} 	
 	
\newcommand{\KS}{\boldsymbol{K}} 
\newcommand{\KT}{\boldsymbol{K}_\perp} 	
\newcommand{\KTS}{K_\perp} 	

\newcommand{\qo}{q_\mathrm{o}} 
\newcommand{\qs}{q_\mathrm{s}} 	
\newcommand{\ql}{q_\mathrm{l}}

\newcommand{\Rl}{R_\mathrm{l}}	
\newcommand{\Ro}{R_\mathrm{o}}	
\newcommand{\Rs}{R_\mathrm{s}}	

\newcommand{\Rol}{R_\mathrm{ol}}	
\newcommand{\Rsl}{R_\mathrm{sl}}	
\newcommand{\Ros}{R_\mathrm{os}}

\newcommand{\GeV}{\,\mathrm{GeV}}
\newcommand{\MeV}{\,\mathrm{MeV}}
\newcommand{\TeV}{\,\mathrm{TeV}}
\newcommand{\fm}{\,\mathrm{fm}}
\begin{document}

\preprint{APS/123-QED}

\title{Probing the evolution of heavy-ion collisions using direct photon interferometry}

\author{Oscar Garcia-Montero$^{a}$}
\author{Nicole L\"oher$^{b}$}
\author{Aleksas Mazeliauskas$^{a}$}
\author{J\"urgen Berges$^{a}$}
\author{Klaus Reygers$^{b}$}
\affiliation{
$^{a}$Institut f\"ur Theoretische Physik, Universit\"at Heidelberg,
Philosophenweg 16, 69120 Heidelberg, Germany \\
$^{b}$Physikalisches Institut, Universit\"at Heidelberg, Im Neuenheimer Feld 226, 69120 Heidelberg, Germany
}

\date{\today}
\begin{abstract}
We investigate the
measurement of Hanbury Brown-Twiss (HBT) photon correlations as an experimental tool to
discriminate different sources of photon enhancement, which are proposed to
simultaneously reproduce the direct photon yield and the azimuthal
anisotropy measured in nuclear collisions at RHIC and the LHC. To showcase
this, we consider two different scenarios in which we enhance the yields from standard hydrodynamical simulations.
In the first, additional photons are produced from the early pre-equilibrium
stage computed from the ``bottom-up" thermalization scenario. In the second, the thermal rates are enhanced close to the
pseudo-critical temperature $T_c\approx 155\,\text{MeV}$ using a phenomenological ansatz.
 We compute the correlators for relative momenta
$\qo, \,\qs$ and $\ql$ for different  transverse pair momenta, $\KTS$, and find
that the longitudinal correlation is the most sensitive to different photon
sources. Our results also demonstrate that including  anisotropic pre-equilibrium
rates enhances non-Gaussianities in the correlators, which can be quantified
using the kurtosis of the correlators. Finally, we study the feasibility of
measuring a direct photon HBT signal in the upcoming high-luminosity LHC runs. 
Considering only statistical uncertainties, we find that with the projected $\sim 10^{10}$ heavy ion events a
measurement of the
HBT correlations for $\KTS<1\, \text{GeV}$ is statistically significant.

\end{abstract}

\pacs{Valid PACS appear here}
\maketitle

\section{\label{sec:Introduction}Introduction}

The relativistic nuclear collision experiments explore the physics of 
dense and hot QCD matter, also known as the quark-gluon plasma (QGP)~\cite{Busza:2018rrf}. The bulk properties
of this new state of matter are inferred indirectly from the yields and correlations
of the produced hadrons. However the QCD degrees of freedom participate in the 
strong interaction and are subject to the effects of multiple-rescatterings and
non-perturbative physics of hadronization, which tend to erase the information about
the earlier stages of the collision.
Electromagnetic probes, e.g.\ photons and dilepton production, are therefore often championed as
penetrating probes of the QGP dynamics~\cite{Chatterjee:2009rs}.
Although it is true that photons escape virtually unscathed from the medium,
the continuous electromagnetic emission makes it very hard to discriminate
between different photon sources. 
Furthermore, in the standard hydrodynamical picture, it is  challenging to
simultaneously describe the measured photon yields and their azimuthal
anisotropy, which is commonly referred to as the  \textit{direct-photon puzzle}
\cite{Adare:2014fwh,Adare:2015lcd,Paquet:2015lta,ALICE,Acharya:2018bdy,David:2019wpt}.

In this paper, we explore two-photon interferometry, called in this
context \textit{femtoscopy}, as a tool to untangle the space-time evolution 
of the QGP and in order to shed light on the direct-photon puzzle.
This addresses the question whether direct photons in heavy-ion collision (HIC)
originate predominantly from the early or the late stage of the collision.
Specifically, we use Hanbury Brown-Twiss (HBT) correlations,
which are the only known way how to directly extract space-time information
from the particles measured in heavy-ion collision experiments~\cite{Heinz:1996bs}. 
HBT correlations, originally introduced to measure the radii of stars
from the incoming photons~\cite{HanburyBrown:1954amm,
HanburyBrown:1956bqd}, have been used
extensively across physics, from atomic gas correlations in cold atom experiments \cite{CAHBT1,CAHBT2}, to
pion interferometry in 
heavy ion collisions
experiments~\cite{Gyulassy:1979yi,Calligarich:1976kp,Pratt:1984su}.
Interferometry of direct photons as a tool to study the space-time
evolution of a heavy--ion collision was theoretically explored by several
authors, see
\cite{Srivastava:1993js,Timmermann:1994kb,Slotta:1996cf,Srivastava:2004xp,Bass:2004de,Peressounko:2003cf,Frodermann:2009nx,Ipp:2012zb}
and references therein. So far only one measurement in Pb--Pb collisions at
$\sqrt{s_\mathrm{NN}} = 17.3\,\mathrm{GeV}$ at the CERN SPS was reported
\cite{Aggarwal:2003zy}. In view of the upcoming high-luminosity runs at the
LHC~\cite{Citron:2018lsq}, we expect further photon measurements at the TeV energy scale and
therefore present theoretical and experimental analysis of  the HBT signal.

In this work we study HBT correlators in different scenarios. First,
we compute the yield and HBT correlators for a hydrodynamically expanding
quark-gluon plasma  and the subsequent hadronic stage using realistic 2+1D event-by-event
simulations of a heavy ion collision.
We then consider two additional sources of
photons, coming from early and late stages of the expansion respectively. At early times 
 we supplement the thermal yield by including a pre-equilibrium
contribution, which was found in previous work to be on par with the
thermal one~\cite{Berges:2017eom, Monnai:2019vup}. Motivated by the idea that thermal rates might be
enhanced  around the pseudo-critical temperature by confining modes during hadronization~\cite{vanHees:2014ida,vanHees:2011vb}, we 
add another source for photons at late times.
We present a detailed analysis of photon HBT signal
sensitivity to different photon sources and make a realistic estimate of
experimental
statistics needed to measure these signals by the ALICE detector.

\section{\label{sec:HBT} HBT Correlations}

Quantum statistical effects can be used to understand the spacetime distribution of particle sources \cite{Heinz:1996bs,Wiedemann:1999qn, Csorgo:2005gd, Lisa:2005dd}. In the context of HIC experiments, we are interested in finding the spatial extension of the photon source in the fireball. For this, we use the normalized HBT correlator, 
\begin{equation}
C(\PS{1},\PS{2}) = \frac{\displaystyle \TPD{1}{2}{}}{\displaystyle \OPD{1}{} \OPD{2}{}}\, ,
\label{eq:C_dist}
\end{equation}
where the numerator is given by the two-photon distribution, which can be expressed in terms of asymptotic states, i.e. creation and annihilation operators of a gauge field
\begin{equation}
\frac{\rmd N}{\rmd^3 p_1 \rmd^3 p_2 } = \, \sum_{\lambda_{1},\lambda_{2}} \langle a^{\dag}_{\PS{1},\lambda_{1}} a^{\dag}_{\PS{2},\lambda_{2}} 
a^{\,}_{\PS{2},\lambda_{2}} a^{\,}_{\PS{1},\lambda_{1}}\rangle \,.
\label{eq:TPD_op}
\end{equation}
Here, $\PS{n}$ and $\lambda_n$ are, respectively, the spatial momenta of the detected photons and polarization mode of the \emph{n}th photon. In a field theoretical language, this can be computed generally from a four-point correlator of gauge fields in momentum space, for equal in- and outgoing momenta.
The denominator is the product of the invariant yields, and can be expressed with asymptotic states as follows,
\begin{equation}
\frac{\rmd N}{\rmd^3 p }=  \sum_{\lambda} \langle a^{\dag}_{\PS{},\lambda} a_{\PS{},\lambda}\rangle \,. 
\label{eq:OPD_op}
\end{equation}

We can further simplify Eq.~\eqref{eq:C_dist} by splitting the four-point function into connected and disconnected parts. The photon fields during a HIC are not expected to   be highly occupied in-medium. This renders the electromagnetic sector to be a dilute gas of particles, for which the photon-photon interaction vertex is very small. In this case we can loose the connected part, and Wick's theorem states that
\begin{align}
\begin{split}
\langle a^{\dag}_{\PS{1},\lambda_{1}} a^{\dag}_{\PS{2},\lambda_{2}} 
a^{\,}_{\PS{2},\lambda_{2}} &a^{\,}_{\PS{1},\lambda_{1}}\rangle \\ \simeq \,&  \langle a^{\dag}_{\PS{1},\lambda_{1}} a_{\PS{1},\lambda_{1}}\rangle \langle a^{\dag}_{\PS{2},\lambda_{2}} a_{\PS{2},\lambda_{2}}\rangle \\
																&+ \langle a^{\dag}_{\PS{1},\lambda_{1}} a_{\PS{2},\lambda_{2}}\rangle \langle a^{\dag}_{\PS{2},\lambda_{2}} a_{\PS{1},\lambda_{1}}\rangle \,.
\end{split}
\end{align}	

From this it can be seen that the two-photon correlator splits into a trivial (diagonal)  and non-trivial (off diagonal) part. It was shown in Ref.~\cite{Heinz:1996bs} that these correlators can be directly related to scalar Wigner density functions $S(x,K)$ (also called emission function in the literature), where the information over polarization of the sources can simply averaged out using the Ward-Takahashi identity. The correlator is found to be
\begin{equation}
C(\QS,\KS) = 1 + \frac{1}{2}\frac{\left| S(\QS,\KS) \right|^2}{S(0, \PS{1}) S(0, \PS{2}) }\,,
\label{eq:CQK}
\end{equation}
where $S(\QS,\KS)$ is the Fourier transform of the emission function,
\begin{equation}
S(\QS,\KS) = \int \rmd^4 x\, e^{\rmi x\cdot q}\, S(x,\KS)\,.
\label{eq:fourierwigner}
\end{equation}
The result is a version the scalar HBT correlator, modified only by a relative degeneracy factor of $1/2$. The new variables, $q =p_1 - p_2$ and $K=(p_1 + p_2)/2$ are the relative and average momenta for two photons, respectively. In what follows, both $S(\QS,\KS)$ and the photon invariant yield, $S(0,\PS{})$,  will be calculated by associating the Wigner function with photon emission rates, that is 

\begin{equation}
{S}(x,\KS)\leftrightarrow E_K\, \frac{\rmd N }{\rmd^4 x \,\rmd^3 K}\,.
\end{equation}
\subsection{Variables and Approximations}
The detected photons are on-shell, and we express the photons four-momenta
\begin{equation}
p^\mu = (p_\perp\cosh y, p_\perp\cos\varphi, p_\perp\sin\varphi, p_\perp\sinh y)
\end{equation}
with rapidity $y$, transverse momentum $p_\perp$  and azimuthal angle $\varphi$.
For the average and relative momentum variables, $q$ and $K$, defined above, we choose a coordinate system such that 
\begin{align}
K^\mu &= (K^0, \KTS, 0, K^z )\nonumber\\
q^\mu &= (q^0, \qo, \qs, \ql)\,,
\end{align}
i.e. $\KS$ lies in the $x-z$ plane, with $z$ being the beam direction. The $q$ components are called the \textit{longitudinal}, \textit{outwards}, and \textit{side} momenta. We can express them using 
\begin{equation}
\begin{split}
\qo &= (\QT\cdot\KT)/\KTS \\
\qs &= \left|\QT - (\QT\cdot\KT)\KT/\KTS  \right|.
\end{split}
\end{equation}

Since both photons are on-shell, both the  pair and relative momenta will be off-shell, and for two identical particles, they satisfy 
\begin{equation}
q_\mu K^\mu = 0  \quad \Rightarrow \quad q^0 = \frac{\QS\cdot \KS}{K^0}.
\label{eq:ort}
\end{equation}

In the case of pion-pion interferometry, two approximations are taken to
further simplify the computation of the HBT correlator. In the literature they
are commonly referred as the \textit{on-shell} and \textit{smoothness}
approximation~\cite{Heinz:1996bs,Frodermann:2009nx}. For the former, the pair momenta itself is taken to be
on-shell, $K^0 \approx|\boldsymbol{K}| +\mathcal{O}(\QS^2)$. In hadron
interferometry, this can be used to good approximation because of the large
masses of hadrons. Even for pions, the subleading terms are suppressed by $E_K$
for all average momenta. In general, in such calculations, if the corrections
are not suppressed by powers of $\QS^2/\KS^2$, they are controlled by the group
velocity $\boldsymbol{\beta} =\KS/K^0$ \cite{Heinz:1996bs,Frodermann:2009nx}.
However, photons are massless, and this expansion will break at
$|\QS|/(2|\KS|)\sim 1$. Unfortunately, in experimentally realizable settings at
the LHC, the direct photon signal is  contaminated by photons from decays,
which form the vast majority of the signal. This leaves direct photons with a
deficiency in statistics. As a consequence, photon pairs cannot be correlated
for infinitesimal $|\QS|$, with reasonable confidence (see Sec.~\ref{sec:experiment}). This means that in general $|\QS|$ and $|\KS|$ will be
on the same order of magnitude. Nonetheless, for a single Gaussian source, the
correlator half-widths can be computed using this approximation without any
problem. For such a source, the correlator will be perfectly Gaussian and the
radii can be directly extracted by fitting the curves \cite{Chapman:1995nz}, or
by computing the curvature of the correlator at $\QS=0$. In the case of direct
photons, we will be having photons from different sources (stages of the
fireball) which will present different scales. Thus, the condition
$|\QS|/(2|\KS|)\ll 1$ cannot be met for all the kinematic regime. Furthermore,
the Wigner function in Eq.~\eqref{eq:fourierwigner} is generally given for any
combination of momenta. The function can be evaluated off-shell
\cite{Frodermann:2009nx,Heinz:1996bs}, and therefore to avoid unexpected
deviations coming from uncontrolled terms we choose to not use it.

The other approximation normally used in the literature is called  the \textit{smoothness approximation}, which consists of  neglecting the $\QS$ dependence in the denominator of Eq.~\eqref{eq:CQK}, via $S(0, \PS{1,2}) \rightarrow S(0, \KS)$. The correlator is given in this limit 
 \begin{equation}
C(\QS,\KS) = 1 + \frac{1}{2}\left|\frac{\tilde{S}(\QS,\KS)}{\tilde{S}(0, \KS) } \right|^2 \equiv 1+\frac{1}{2}\left\langle e^{i q\cdot x} \right\rangle \,,
\label{eq:CQKSmooth}
\end{equation}
for which we will introduce the commonly used averaging notation~\cite{Frodermann:2009nx}: 
\begin{equation}
\langle f(x) \rangle = \frac{\displaystyle \int \rmd^4 x f(x)\,S(x,K) }{\displaystyle \int \rmd^4 x S(x,K) }\,.
\end{equation}

The smoothness approximation is accurate if the curvature logarithm of the one particle distribution is small \cite{Chapman:1994ax}, which is not true for photons at small momentum (see Fig.~\ref{fig:Yields}). In reference \cite{Frodermann:2009nx} it was found that the convergence of the smoothness approximated to the full correlator is restricted for values of $|\QS|< 2\,|\KS|$. This is the same scale that signals the breakdown of the on-shell approximation. We use this approximation for the case of 1D slices for $\qo$ and $\qs$, where the other directions of $q$ are set to zero. In this case, the $\qo$ and $\qs$ direction look Gaussian, and the explored $\KTS$ values will be larger than the inverse half width of the correlator, which makes this approximation safe. 

\subsection{Homogeneity Radii}
\label{sec:hr}
We can get a general form of $C(\QS,\KS)$ for an arbitrary Gaussian source around the origin in $\QS$ space
\begin{equation}
C(\QS,\KS) =1 + \frac{1}{2} \exp\left[-q_\mu\, \tilde{R}^{\mu\nu}\, q_\nu\right] \,,
\end{equation} 
For sources with relatively small non-Gaussianities this approximation is still valid, since the perturbations around $C$ increase only at high $\QS$ values \cite{Heinz:1996qu}. 
The half-width tensor, $\tilde{R}^{\mu\nu} = \tilde{R}^{\mu\nu}(\KS)$, is a function of the pair momentum. To clean the notation, we avoid writing its $\KS$ dependence. Using the orthogonality relationship, Eq.~\eqref{eq:ort}, we can reduce this expression to 
\begin{equation}
C(\QS,\KS) =1 + \frac{1}{2} \exp\left[-q_i\, R^{ij} \,q_j \right]\,,
\label{eq:cartPar}
\end{equation}
 by redefining $R^{ij} \equiv \beta^i\, \beta^j\,R^{00} +2\, \beta^i\,R^{0j} +R^{ij}$.  Because of symmetry, $R^{ij} =R^{ji}$, we only get 6 independent components. Using the relative momentum parametrization introduced above, we can express it as 
\begin{equation}
R_{ij}(\KS)=
\begin{bmatrix} 
 \Ro^2 & \Ros^2 & \Rol^2 \\
 \Ros^2 & \Rs^2 & \Rsl^2\\ 
 \Rol^2&  \Rsl^2 & \Rl^2
\end{bmatrix}\,.
\label{eq:R}
\end{equation}

In this work we only focus on the diagonal of this matrix. While it has been shown that for longitudinally expanding sources the $\Ros$ term is relevant \cite{Chapman:1994yv}, it is also true that numerically calculating such cross-terms is more computationally complex.  

To compute the radii in Eq.~\eqref{eq:R}, we use the the method of moments, which is stable for correlators with strong non-Gaussianities \cite{Wiedemann:1999qn}. We use the moments of the true correlator $C(\QS, \KS)-1$ in relative momentum space, 
\begin{equation}
\langle \langle q_i q_j \rangle \rangle  = \int \rmd^3q\, q_i\, q_j\,  g(\QS;\KS)\equiv \frac{1}{2}(R^{-1})_{ij}\,,
\end{equation}
where $R^{-1}$ is the inverse matrix of Eq.~\eqref{eq:R}.  We have defined the  distribution function 
\begin{equation}
g(\QS;\KS) \equiv \frac{C(\QS,\KS)-1}{\int \rmd^3 q\, \left[C(\QS,\KS) -1\right]}
\end{equation}
to ensure correct normalization. Because of the symmetry properties of the correlator, we can safely assume the one-point functions vanish, $\langle \langle q_i \rangle \rangle = 0$. For simplicity, and because we do not explore the off-diagonals, we will keep the notation one-dimensional.  That means that the homogeneity radii are going to be given by 
\begin{equation}
R_i^2 = \frac{1}{2} \langle \langle q_i^2 \rangle \rangle^{-1} , \quad \text{with} \quad i\in \{\mathrm{l,o,s}\}.
\end{equation}

It is important to clarify that this method requires the correlator to be highly localized around $\QS=0$, to give sensible results for the characteristic scale. In other words, the correlator needs to decay faster than a power-law. We can use also this method to quantify the deviations from Gaussianity by computing the \textit{normalized  excess kurtosis}, 
\begin{equation}
\Delta_{i} = \frac{\langle\langle q^4_i \rangle\rangle}{3\langle\langle q^2_i\rangle\rangle^2 }-1,
\label{eq:NEK}
\end{equation}
which, as expected, vanishes in the Gaussian limit. In theoretical calculations of HBT correlations, going to higher values of $q_i$ requires only better numerical precision. However, it may be problematic for experiment, where high relative momentum values will suffer from statistic limitations.

\section{\label{sec:model} Modelling the photon sources}
As it was stated in the introduction, we calculate the thermal photon observables, which are enhanced by the inclusion of early- and late-time photon sources. The thermal base is calculated from hydrodynamic simulation using the VISHNU package~\cite{VISHNU,VISHNUwebsite,Shen:2014vra}, from which realistic space-time evolution of temperature and velocity fields was obtained. Using the default model parameters tuned to the experimental data, we simulated 200 Pb-Pb collision events  at the centre of mass energy
 $\sqrt{s_\text{NN}}=2.76\TeV$ in 0-20\% centrality class. The initial conditions at $\tau_{\mathrm{hydro}}=0.6\fm$ were provided by the two-component Monte Carlo Glauber model~\cite{Shen:2014lye}. The relativistic hydrodynamic simulation was then performed using fixed shear viscosity over entropy ratio $\eta/s=0.08$ and the decoupling energy density $e=0.1\GeV/\text{fm}^3$.
 The space time evolution of transverse velocities $v_x$ and $v_y$ and temperature $T$ was recorded on
 a coarsened grid with spacing $dx=dy=0.4\fm$ and $d\tau=0.2\fm$ ($x_\text{max}=y_\text{max}=25.2\fm$). The final time $\tau_\text{max}$ varied depending on the initial conditions, but at least 100 recorded events had $\tau_\text{max}\geq 15.8\fm$. We calculate photon emission for each event separately and then do the ensemble average.
 
Direct photons can be emitted from the QGP and hadron resonance gas (HRG) epochs of the evolution of the fireball. The transition from the QGP production to the HRG is signaled by a switch at 160 MeV. It is assumed that the emission threshold for thermal photons is at a temperature of 120 MeV. In addition, two possible sources for enhancing the invariant photon yield are discussed. The first is the inclusion of a pre-equilibrium source based on the first stage of the \textit{bottom-up} thermalization scenario \cite{Baier:2000sb, Berges:2017eom}. The second source is a phenomenological enhancement of the thermal rates near a pseudo-critical temperature $T_{pc}$, presented first in Ref. \cite{vanHees:2014ida}. We discuss these  and other photon contributions below.

\subsection{\label{sec:prompt}Prompt photons from the initial stage}

During the initial stage of the collision, \textit{prompt} photons are produced via hard scattering of the partons from the individual nucleons. The photon cross-section for the $NN \rightarrow \gamma X$ process can be calculated using perturbative QCD (pQCD) \cite{VogelsangPrivate}, which is then scaled by the number of binary collisions,  $N_{\mathrm{coll}}$, via the relation

 \begin{equation}
\frac{\mathrm{d} N_{\mathrm{prompt}}}{\mathrm{d}^2 p_\perp\mathrm{d}y}= \frac{N_{\mathrm{coll}}}{\sigma^{\mathrm{NN}}_{\mathrm{inel}}}\, \frac{\mathrm{d}\sigma^{NN \rightarrow \gamma X} }{\mathrm{d}^2 p_\perp\mathrm{d}y}\,.
\label{eq:totaldirect}
\end{equation}
Here $\sigma^{\mathrm{NN}}_{\mathrm{inel}}$ is the total inelastic collision for a collision of two nucleons. We compute $N_{\mathrm{coll}}$ using the optical Glauber model. For the computation of the full photon invariant yield we need to extend the pQCD computation to smaller $p_\perp$ values. We do so by taking the same parametrization used by PHENIX Ref.~\cite{Adare:2014fwh}. The fit function is given by the functional form 
\begin{equation}
 \frac{\mathrm{d}\sigma^{pp} }{\mathrm{d}^2 p_\perp\mathrm{d}y} =  A_{pp}\,\left(1+\frac{p^2_\perp}{P_0}\right)^{-n}\,.
\end{equation}
Because this contribution takes on account incoherent production of single photons, we do not include prompt photons in the calculation of the HBT correlator, but add them to the total photon yield.

\subsection{\label{sec:AMY}Photon emission from the quark-gluon plasma}

To compute the photon contribution due to the thermal QGP we use the full leading order (LO) computation, parametrized in Ref.~\cite{Arnold:2001ms}. This rate contains not only the two-to-two contributions which dominate at higher momenta, but also near-collinear bremsstrahlung and the inelastic pair annihilation, thereby fully including the Landau-Pomeranchiuk-Migdal effect (LPM), which can be understood as suppression of emission owing to interference of multiple scatterings \cite{Aurenche:2000gf,AURENCHE1,AURENCHE2}. The parametrization used in this work is given explicitely in Appendix \ref{app:rates}.

\subsection{\label{sec:RAPP}Photon emission from the hadron resonance gas}

For the thermal photon emission rate from the hadron resonance gas phase the parametrizations of Ref.~\cite{Heffernan:2014mla} is used. The given parametrizations agree within 20\% with the microscopic calculated values. Microscopic calculations have already been performed~\cite{Rapp:1999us,Liu:2007zzw}, but, as pointed out in~\cite{Heffernan:2014mla}, the results cannot be easily used in models like the one described here. Two different parametrizations for the photon emission rate are given: one for the contribution from the in-medium $\rho$ mesons and one for the contribution from bremsstrahlung originating from $\pi\pi$ scattering. They can be applied to photons with energies $q_{0}$ between 0.2 and 5 GeV, which are produced from chemically equilibrated matter with a temperature between 100 and 180 MeV and baryon chemical potentials of 0 to 400 MeV. In the case of ALICE, vanishing chemical potential is assumed.

\begin{figure*}
\includegraphics[width = 0.8\textwidth]{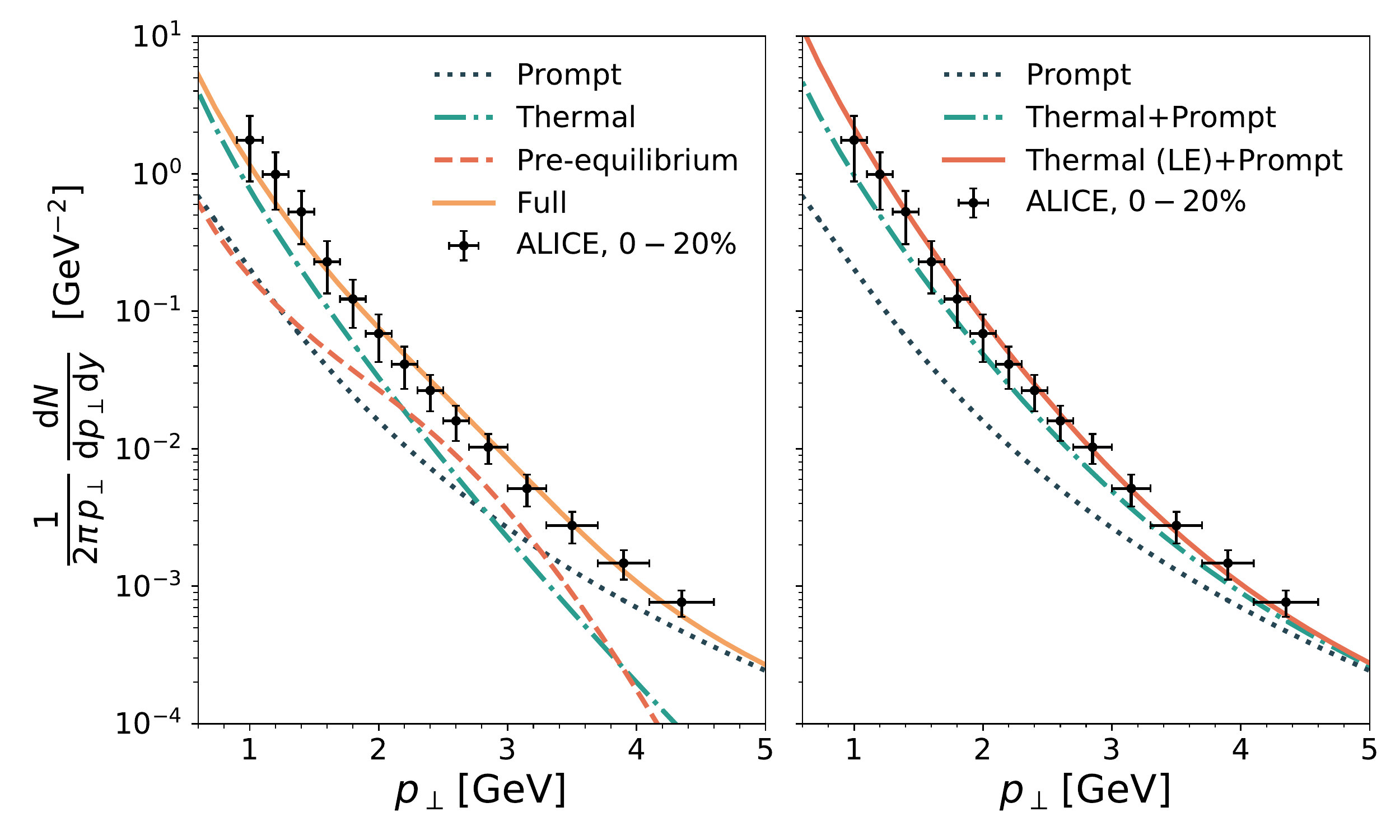}
\caption{\label{fig:Yields} 
Comparison of invariant yield of direct photons from different sources to ALICE measurement in
central (0-20\%) Pb--Pb collisions at $\sqrt{s_\mathrm{NN}} =
2.76\TeV$~\cite{ALICE}.
On the left we show photon contributions from prompt (dotted line), thermal (dash-dotted line) and early-time pre-equilibrium (dashed line) sources. The total result is shown by the solid line.
On the right we compare the combined prompt and thermal photon yield (dash-dotted line) with 
the late-time pseudo-critical enhancement (LE)
scenario (solid line).
}
\end{figure*}

\subsection{\label{sec:CYM} Photon production from pre-equilibrium}

Using the  ``bottom-up" thermalization scenario~\cite{Baier:2000sb}, recent estimates \cite{Berges:2017eom,Khachatryan:2018ori} show the pre-equilibrium contribution photons to be non-negligible. The central idea in this estimate is that gluon saturation takes place at RHIC and LHC energies, which means that during the initial stage of a collision, the nuclei behave as macroscopic fields, and undergo very strong, non-linear interactions. After a parametrically short time given by the saturation scale, $\tau_0\sim Q_s^{-1}$, the gluon fields get highly occupied and undergo three stages of relaxation. During the first stage, hard modes with $p_\perp \sim Q_s$ completely dominate the system. These modes are approximately conserved, yet diluted thanks to Bjorken expansion. During this stage, hard modes scatter via two-to-two scatterings, which produces a broadening of the distribution in the $p_z$ direction. 
The second stage starts once the occupation of the gluon modes falls below unity, where the typical longitudinal momentum of hard gluons saturates at a finite value. In this stage, hard gluons still dominate the total gluon number, while the typical interactions are taken over by the soft sector. Finally, we arrive to the third stage of the BMSS scenario, where the number of soft and hard gluons becomes comparable. Soft gluons thermalize rapidly via two-to-two scatterings, which creates a bath to which hard gluons quickly loose energy to, via mini-jet quenching. The system has then fully thermalized. 

We fix the initial characteristic scale IP-Glasma model \cite{Schenke:2012wb}, which combines the
geometry of the MC-Glauber model \cite{Miller:2007ri} with the IP-Sat model
\cite{Kowalski:2003hm, Rezaeian:2012ji}, while the BMSS scenario gives the time
dependence of the rates. We use as well experimental data to constraint the
needed parameters, the thermalization time was found in Ref.
\cite{Berges:2017eom} to be $\tau_\mathrm{th}\sim 2.4\fm$ for LHC and RHIC
energies. Since the \textit{bottom-up} scenario does not account for the
transverse expansion , such late thermalization poses a phenomenological
problem, as the photons will not be able to build up enough anisotropy,
creating tension with data. To avert this, we will only evolve the
pre-equilibrium stage up to the end of the first stage of the
\textit{bottom-up} scenario, $\tau_{\mathrm{hydro}}=0.6\fm$
\footnote{The photon spectra from all three stages of the \textit{bottom-up} thermalization is studied in Ref.~\cite{Garcia-Montero:2019vju}}.
From the field
theoretical point of view, in this stage, the gluon medium approaches a
non-thermal fixed point \cite{Berges:2013eia,Berges:2013lsa,Berges:2013fga},
where the gluon occupation is given by
\begin{equation}
f_g(\tau; \,p_\perp, p_z)= \frac{1}{\alpha_S}\left(\frac{\tau}{\tau_0}\right)^{-\frac{2}{3}}\,f_S \left(p_\perp,  p_z\,\left(\frac{\tau}{\tau_0}\right)^{\frac{1}{3}}\right).
\label{eq:glue1}
\end{equation}
Here, $\alpha_S$ is the strong coupling, and $f_S$ is a scaling function, which can be parametrized from the results of classical statistical simulations \cite{Berges:2013eia} as follows
\begin{equation}
f_S(p_\perp, p_z)=f_0 \frac{Q_s}{p_\perp}\, \exp\left[{-\frac{1}{2}\frac{p_z^2}{\sigma^2_0}}\right]\, W_r[p_\perp-Q_s]\,.
\label{eq:glueS}
\end{equation}
Here, $W_r[p_\perp,Q_s]$ stands for a suppression function, inspired by the classical statistical simulations. It depends on a free suppression parameter, $r$, and it is given by 
\begin{equation}
\begin{split}
W_r[p_\perp,Q_s]=& \, \theta(Q_s - p_\perp)\\&+  \theta(p_\perp-Q_s) 
\,e^{-\frac{1}{2}\left(\frac{p_\perp-Q_s }{r\,Q_s}\right)^2}\,.
\end{split}
\end{equation}

At the end of this stage, the system is assumed to instantaneously thermalize, and we match the energy densities in the pre-equilbrium  and hydro stages at $\tau_{\mathrm{hydro}}$, 
\begin{equation}
\epsilon_{\mathrm{early}}(\tau_{\mathrm{hydro}},\boldsymbol{x}_\perp) = \epsilon_{\mathrm{hydro}}(\tau_{\mathrm{hydro}},\boldsymbol{x}_\perp)\,, 
\end{equation}
which gives also the spatial profile of the saturation scale $Q_s(\boldsymbol{x}_\perp)$. 
In the pre-equilibrium stage, most of the energy density resides in the gluonic sector. Using Eq.~\eqref{eq:glue1} and the QGP energy density one can obtain
\begin{equation}
\frac{Q_s(\boldsymbol{x}_\perp)}{T(\boldsymbol{x}_\perp)}= \left[\sqrt{\frac{2}{\pi}}\frac{37\pi^2(2 \pi)^2\alpha_S}{30(1 + \sqrt{2\pi}\,r\, + 2\,r^2))} \frac{\tau_{\mathrm{hydro}}}{\tau_{0}f_0\,s}  \right]^{1/4}
\end{equation}
where $\tau_0$ is taken to be the spatially averaged saturation scale, $\langle Q_s\rangle$, and can be determined parametrically using the method described in Ref.~\cite{Berges:2017eom}.

For the rate, we will use a kinetic rate, generally given by 
\begin{equation}
\begin{aligned}
E\frac{\mathrm{d}N }{\mathrm{d}^4X \mathrm{d}^3p }\, =\, & \frac{1}{2\,(2\,\pi)^{12}}\int \frac{\mathrm{d}^3\,p_3}{2E_3} \frac{\mathrm{d}^3\,p_2}{2E_2}  \frac{\mathrm{d}^3\,p_1}{2E_1}\,|\mathcal{M}|^2\\
&\quad \times\, (2\,\pi)^4\, \delta^4(P_1+P_2-P_3-P) \\
& \quad\times\, f_1(p_1)\,f_2(p_2) \left[1\pm f_3(p_3)\right]\,, 
\end{aligned}
\label{eq:kinetic}
\end{equation}
where the processes included are the two-to-two annihilation, $qq\rightarrow g\gamma$, and Compton scattering, $qg\rightarrow q\gamma$. Because the computation at each space-time point of such rate requires a 5-dimensional integral, we simplify the rate using the small angle approximation.  For massless mediators, hard scatterings present collinear enhancement, which will dominate the integrals in Eq.~\eqref{eq:kinetic}. Expanding in the exchange momentum of the mediator and keeping only the leading term one finds the simplified rate \cite{Berges:2017eom, Blaizot:2014jna}, 
\begin{equation}
E\frac{\mathrm{d}N }{\mathrm{d}^4\,x \mathrm{d}^3\, p }= \frac{10}{9\pi^4} \, \alpha \,\mathcal{L}\,Q^2_s \, \kappa_g\, \left(\frac{\tau_0}{\tau}\right)  f_q (\tau,\mathbf{p})
\label{eq:rate}
\end{equation}
where $\alpha$ is the electromagnetic coupling, $\kappa_g=c\,(2N_c)^{-1}$, where $N_c$ is the number of colors, and $c$ is the gluon liberation factor described in Ref.~\cite{Lappi:2007ku}. The quark distribution, $f_q$ is taken from hard splitting of gluons in-medium, namely  $f_q\sim \alpha_S\,  f_g$. That is, using this parametrization, we assume the quark distribution inherits the scaling properties of the parent gluons. To avoid breaking fermion statistics, we suppress the quark distribution for low $p_\perp$ values, so that $f_q=1/2$ at its highest value. The $\mathcal{L}$ term is called the Coulomb logarithm, and it is a regulator, which relates the UV and IR scales, two cutoffs which are needed for this approximation. In the thermal case, the UV scale can be related to the temperature, $T$, while the IR scale can be related to the Debye mass, $m_D \sim gT$. Using this identification, the leading-log (LL) thermal rate from Ref.~\cite{KAPUSTA} can be found from the small-angle approximated rate. 

Nevertheless, at the full leading-order (LO) limit of the photon rate, Ref.~\cite{Arnold:2001ms}, it was shown that in a thermal setting, photon rates are dominated by near-collinear bremsstrahlung for photon energies or $p\lesssim 2\,T$, while at $2\,T\lesssim p \lesssim 10\, T$, the two-to-two terms are of the same order to the near-collinear contributions. The modification for the rate is applied then by changing the constant under the log
\begin{equation}
\mathcal{L}\rightarrow \nu_{LO}(x)
\label{eq:NULO}
\end{equation}
where $x =E/T$ in the thermal case, and $\nu_{LO}(x)$ is given in Eq.~\eqref{eq:L_LO}. We expect a similar behavior to the pre-equilibrium stage, with one difference. During this stage, the characteristic momentum scale is taken to be the saturation scale $Q_s$, making the near-collinear contributions during the early stages dominant at $p\lesssim 2\,Q_s$ which for the center of mass energy at ALICE is most of the kinematic window at which direct photons are observed. We therefore also use the modification of Eq.~\eqref{eq:NULO} in Eq.~\eqref{eq:rate}, for $x\rightarrow x'=E/Q_s$.

\subsection{\label{sec:TC} Critical enhancement at late times near $T_{c}$}

To account for the missing photons one could naively push the initial time to smaller values. Nevertheless, doing so hardens the spectrum, which creates tension with the experimental results \cite{vanHees:2014ida,Paquet:2017wji}. If one has to increase the thermal rate, it has to be done increasing the weight of photons coming from later times.This is in line with the idea suggested in Refs.~\cite{vanHees:2014ida,Rapp:2013ema,vanHees:2011vb,Shen:2013vja}, where it is conjectured that the thermal rates are enhanced near a pseudo-critical temperature $T_{\mathrm{c}}$, 
\begin{equation}
E \frac{d N_{\mathrm{enh}}}{\rmd^4 x\, \rmd^3 p} \equiv  h(T)\, E \frac{d N_{\mathrm{thermal}}}{\rmd^4 x\, \rmd^3 p}
\label{eq:enh1}
\end{equation}
by the fact that close to the transition to hadronic degrees of freedom, one has to account for interactions related to confinement. This means that the partonic cross-sections will see a rise which cannot be accounted for by perturbative physics \cite{Kaczmarek:2005ui}. 
For the purpose of this paper, however, we choose to model the enhancement factor, $h(T)$, as follows
\begin{equation}
\begin{split}
h(T) &=1+h_0\, e^{-\frac{(T-T_\mathrm{c})^2}{d^2}}
\end{split}
\label{eq:enh2}
\end{equation}
where the pseudo-critical temperature is set to be $T_{\mathrm{c}}=155\MeV$. The enhancement parameters are set to be $h_0= 3 $ and $d = 50\MeV$. The enhancement factor is tuned such that the enhancement matches the experimental results from the ALICE collaboration, see Fig.~\ref{fig:Yields}.

\section{\label{sec:results}  Results}

\begin{figure}
\includegraphics[scale= 0.47]{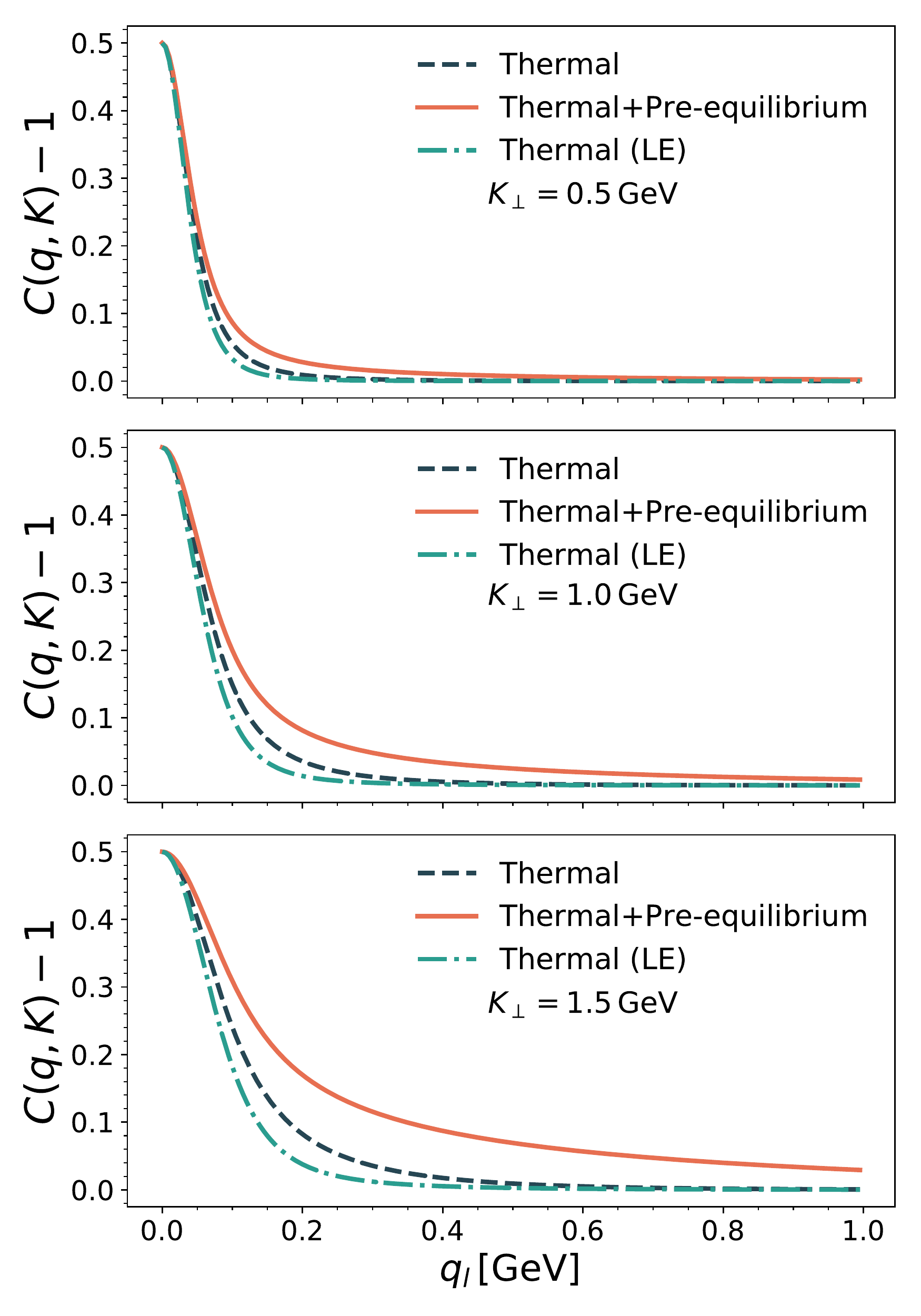}
\caption{\label{fig:CL_HM} The HBT signal for $\ql$ and $\qs=\qo=0 $ for  thermal (dashed line), thermal and pre-equilibrium (solid line) and  thermal photons with late enhancement (dash-dotted), for $\KTS =0.5,1.0,1.5\GeV$. Deviation from the thermal stage can be seen increasing with $\KTS$ for both enhancement scenarios.}
\end{figure}

We compute the total yield for the thermal baseline, and include as well the both enhancement scenarios, which can be seen in Fig.~\ref{fig:Yields}.  The pre-equilibrium photon spectrum shows a structure around $p_\perp \sim 2.5\GeV$. This shoulder comes directly from the parametrization of the quark function. Nevertheless, the specific value at which we can find the peak is given by averaging the space dependence of $Q_s(\boldsymbol{x}_\perp)$. The pre-equilibrium spectrum is found to be dominant for $2\GeV < p_\perp \sim 3\GeV$, while being relatively small in the IR sector. Summing over the prompt, pre-equilibrium and thermal contributions we find good agreement with ALICE data for central collisions, $0-20\%$ (Fig.~\ref{fig:Yields}, left).
On the other hand, applying the enhancement to the thermal rates, Eqs.~\eqref{eq:enh1} and  \eqref{eq:enh2},  just as expected, we see an overall increase of the spectrum, particularly strong for low-$p_\perp$, photons. It can be seen that both scenarios are compatible with the errorbars, which means that distinguishing such cases experimentally is not possible using only the invariant yield. 

\begin{figure}
\includegraphics[scale= 0.425]{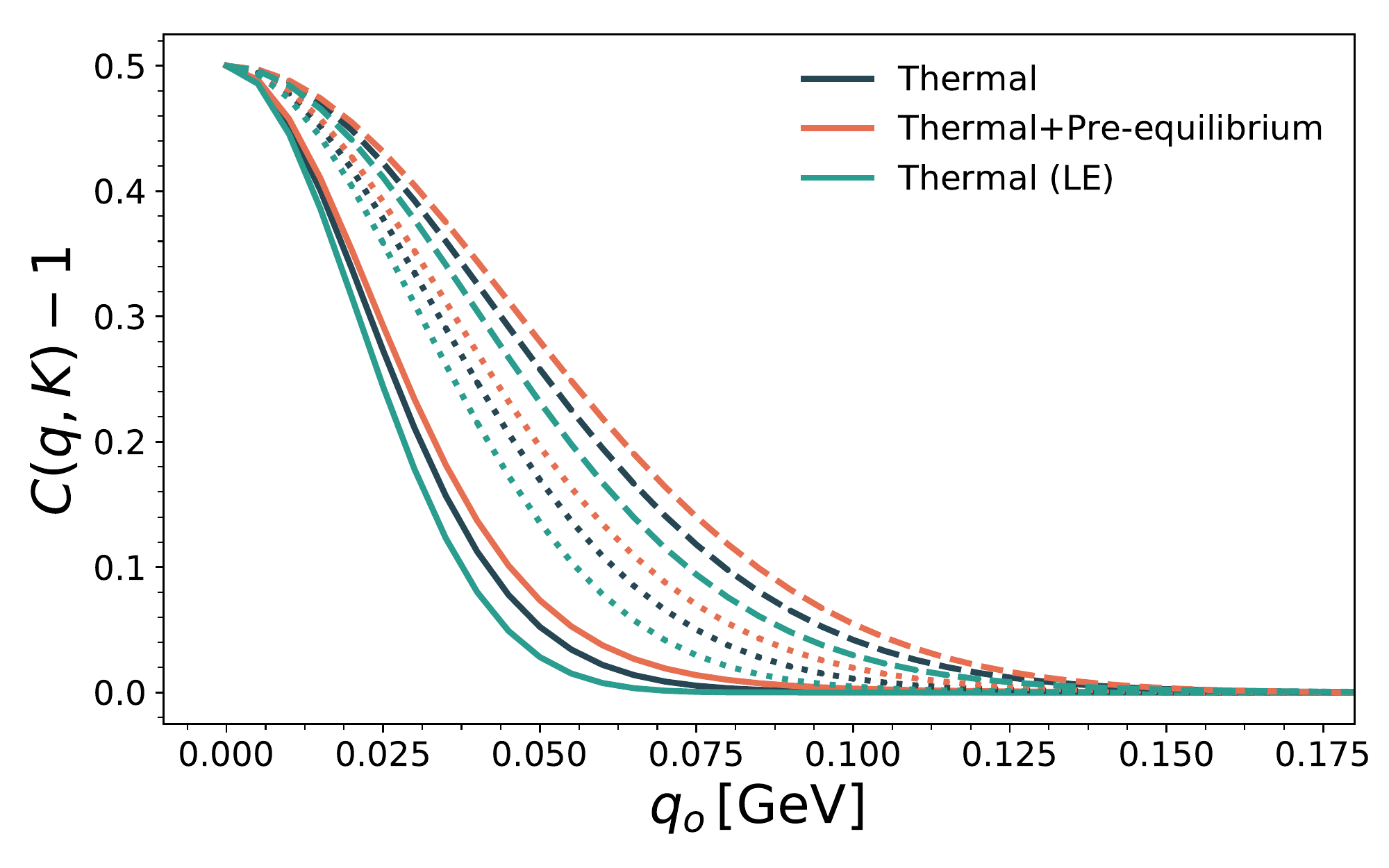}
\includegraphics[scale= 0.425]{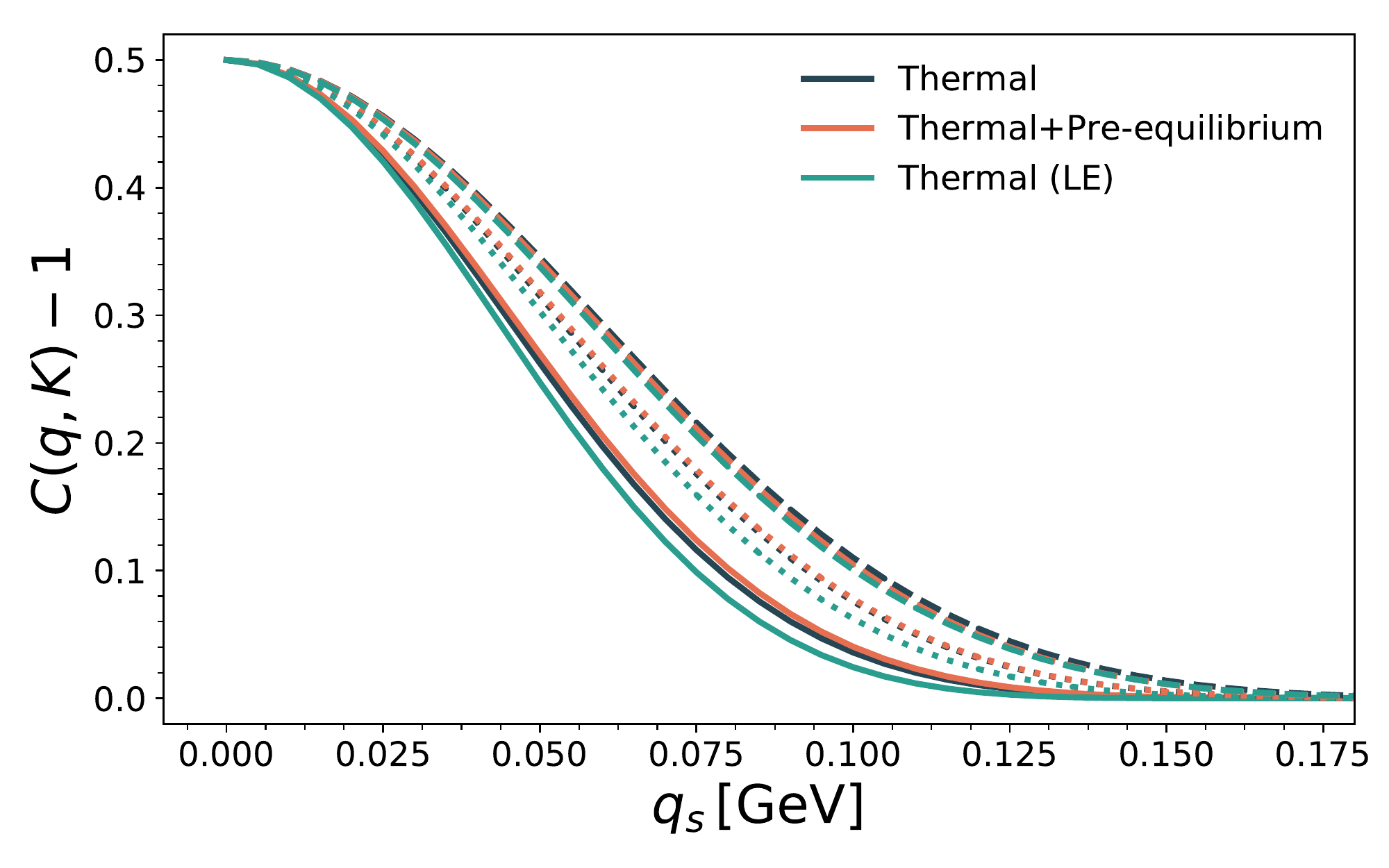}
\caption{\label{fig:OSCorrs} (top) Outward and (bottom) sideward correlators, for $K_\perp=0.5\GeV$ (solid lines), $K_\perp=1.0\GeV$ (dotted lines) and $K_\perp=1.5\GeV$ (dashed lines).  }
\end{figure}

\begin{figure*}
\includegraphics[scale= 0.4]{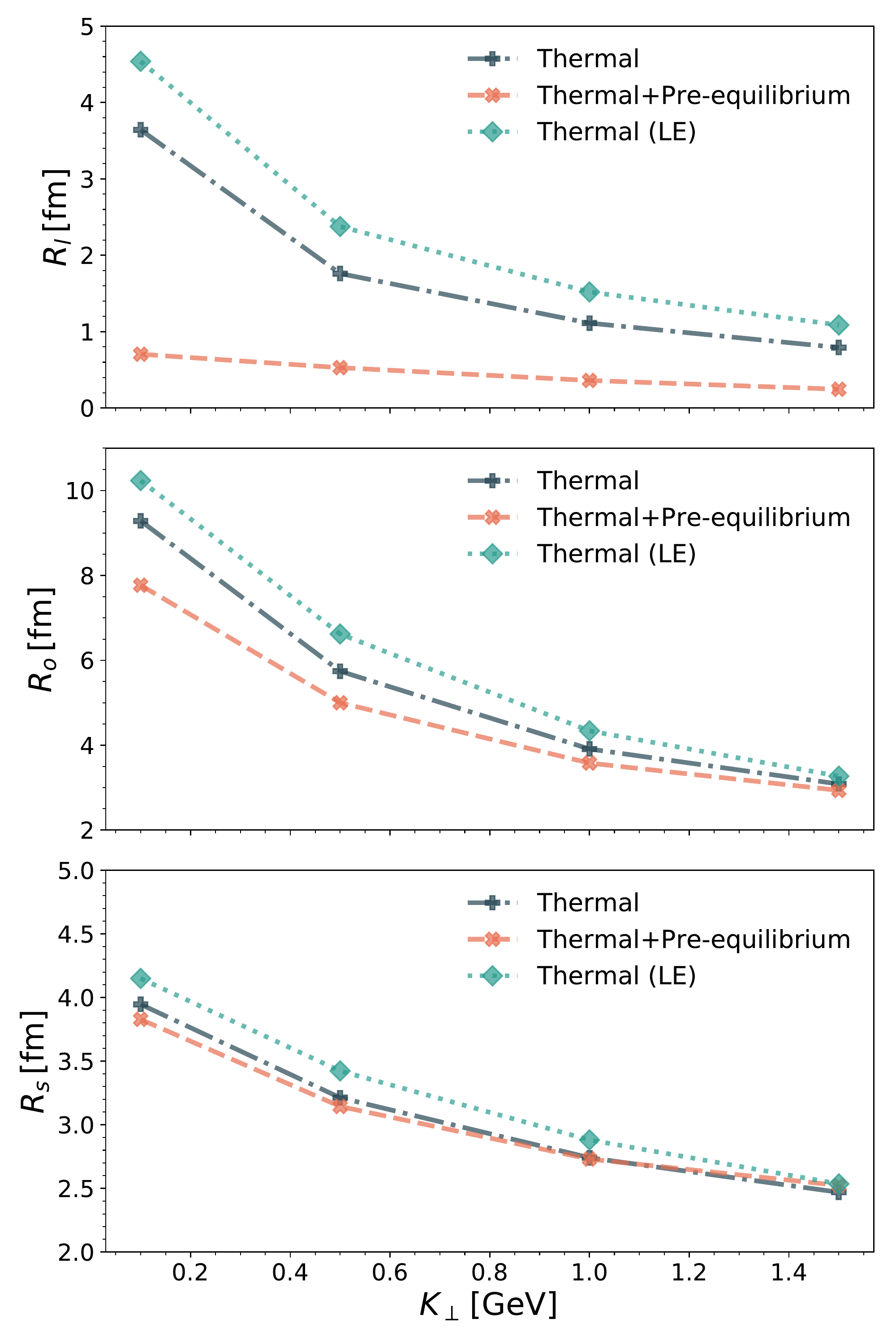}
\includegraphics[scale= 0.4]{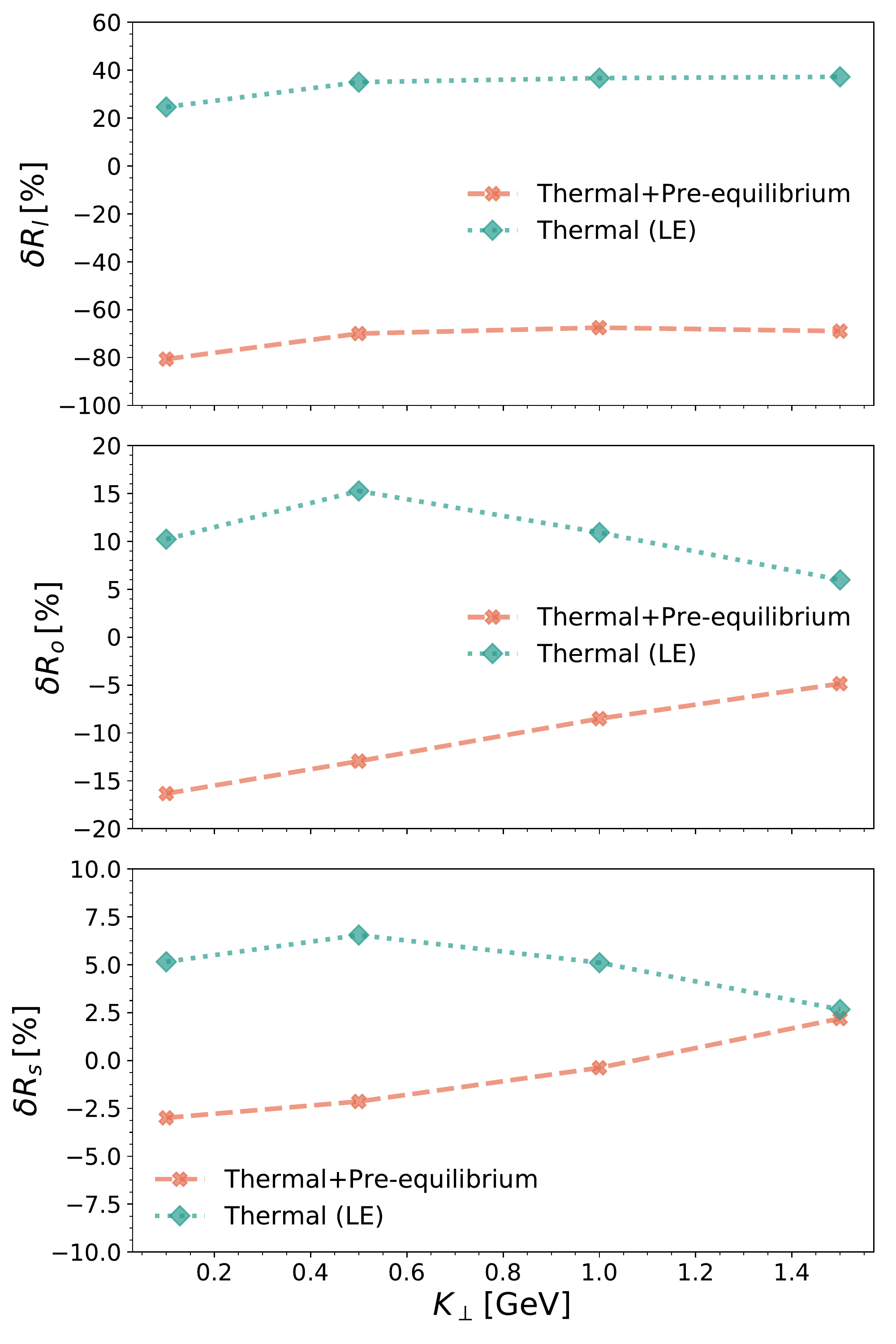}
\caption{\label{fig:RLOS} (left) HBT radii as a function of the pair momenta, calculated using the method of moments. (right) Percentage change of the radii for the two scenarios. }
\end{figure*}

The full HBT correlator, Eq.~\eqref{eq:CQK}, was computed for midrapidity pairs, $K_z=0$, along the three diagonals, i.e. $q_i$ with $q_j=q_k=0$ for $i\neq j\neq k$. We focus on $0-20\%$ central collisions in ALICE, with $\sqrt{s_\mathrm{NN}}=2.76\TeV$, where the average saturation scale is $\langle Q_{s}^2\rangle = 2.9\GeV^2$. 
As expected, the longitudinal curves are the most sensitive to the inclusion of both enhancements which are presented in Fig.~\ref{fig:CL_HM} for different values of $\KTS$. Although the correlator around the side- and outward diagonals show a difference with the inclusion of both enhancements, the effect is noticeably small. This can be seen better for the diagonal radii,  $\Rl$, $\Ro$ and $\Rs$ (see Fig.~\ref{fig:RLOS}), which were computed using the characteristic scale method and the aforementioned correlators. 

 Just as expected from the correlators, the change in the longitudinal radius, $\Rl$, is the largest one. The change induced in $\Ros$ by the inclusion of the scenarios was found it to be in the $10-20\%$ range for the outward direction, and $0-10\%$ for the sideward direction. The small change in the transverse radii will make using them to discriminate models difficult. Nevertheless, this gives an interesting case for predictions. Take, for example, the pre-equilibrium case: If pre-equilibrium photons are relevant at the yield level, and the assumption that the pre-equilibrium stage does not create enough pressure gradients is correct, thermal models will be able to reproduce the $\Ros$ but may undershoot significantly $\Rl$. On the other hand, a consistent increase with $K_\perp$ on the three radii may indicate that photons come from the late stages.

\begin{figure}
\includegraphics[scale= 0.35]{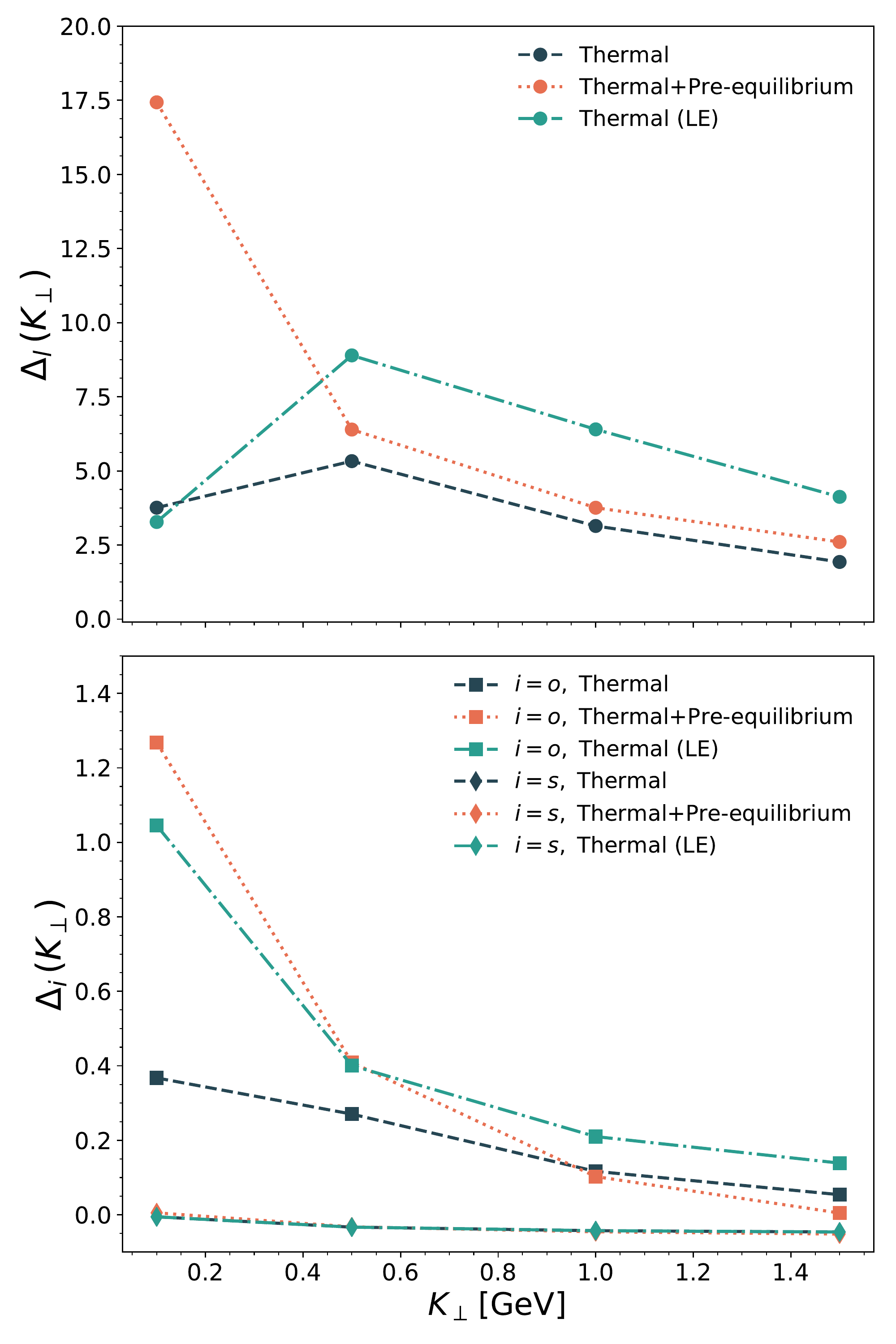}
\caption{\label{fig:NEKS}Normalized excess kurtosis for the $q_i$ direction, with $i=\mathrm{l,o,s}$. The strongest difference from Gaussianity is seen in the longitudinal direction, followed by the outwards direction. Finally, the sidewards direction is approximately Gaussian.}
\end{figure}

We  also computed the normalized excess kurtosis, Eq.~\eqref{eq:NEK},  for the three diagonals. A clear hierarchy is found, where $\ql$ breaks Gaussianity the most, followed by $\qo$ and $\qs$. We find that the sidewards direction is to good approximation Gaussian (see Fig.~\ref{fig:NEKS}). The non-Gaussianities, as was explained above and in Ref.~\cite{Frodermann:2009nx} arise from the longitudinal expansion of the fireball. In the case of massless particles these effects will be considerable more important than for e.g. pions. Additionally volume emission will further enhance these effects, opposed to Cooper-Fry surface emission. Non-Gaussianities are quite intuitive to understand in the case of the $\ql$ direction, since the boosting from longitudinal expansion is largest for the $\ql$ variable. However, the easiest way to see how the outward direction gets contributions from the expansion is the definition $R^{ij} \equiv \beta^i\, \beta^j\,R^{00} +2\, \beta^i\,R^{0j} +R^{ij}$. From this  formula we see that for the outward direction, $\Ro$ gets a non zero contribution from $\beta_o t =t  K_\perp/K^0 $, while the sideward direction, by the definition, will not. This means that the outward homogeneity radius not only depends on the spatial size of the source, but also on the lifetime of emissions \cite{Heinz:1996bs}. As it can be seen in Fig.\ref{fig:NEKS}, the normalized excess kurtosis can be used as an observable complementary to the radii. This is particularly true for $K_\perp < 0.5\GeV $, where the big difference in $\Delta_l$ could be used to differentiate the scenarios.

\section{\label{sec:experiment} Experimental feasibility}

Measuring direct-photon Hanbury Brown-Twiss correlation is a challenging task. At the LHC, the ALICE experiment measures photons at low transverse momentum ($\lesssim 3\,$GeV) \cite{ALICE,Acharya:2018bdy}. Significantly improved data-taking rates in the upcoming LHC runs 3 and 4 make it possible to collect a sample of Pb--Pb collisions corresponding to an integrated luminosity of $10\,\mathrm{nb}^{-1}$, or $\mathcal O (10^{10})$ collision events. In this section we estimate up to what photon pair transverse momentum $\KTS$ a direct-photon HBT measurement might be possible. 

We concentrate on the longitudinal momentum difference $\ql$. Statistical uncertainties for measurements of $\qo$ and $\qs$ are very similar. For a Gaussian parameterization the correlation function $C$ of direct photons for $\qo = \qs = 0$ is given by
\begin{equation}
C = 1 + \lambda \exp(- \Rl^2 \cdot \ql^2)
\label{eq:corr_func_gauss}
\end{equation} 
with $\lambda = 1/2$. The total number of photons, however, is dominated by photons from neutral pion and eta meson decays. Owing to the long lifetime of the neutral pion and the eta mesons the decay photons are not correlated with the direct photons and dilute the measured correlation function, resulting in
\begin{equation}
\lambda = \frac{1}{2} r_\gamma^2, \quad r_\gamma = \frac{N_\mathrm{dir}}{N_\mathrm{inc}}
\end{equation} 
for the correlation strength of pairs of inclusive photons. Here $N_\mathrm{dir}$ denotes the number of direct photons and $N_\mathrm{inc}$ the number of inclusive photons, i.e., the sum $N_\mathrm{inc} = N_\mathrm{dir} + N_\mathrm{dec}$ of the number of direct and decay photons. We assume a $p_\perp$-independent fraction of direct photons of $N_\mathrm{dir}$/$N_\mathrm{inc} \approx 0.1$ corresponding to $\lambda = 0.005$ \cite{ALICE}.

\begin{figure}
\includegraphics[scale= 0.45]{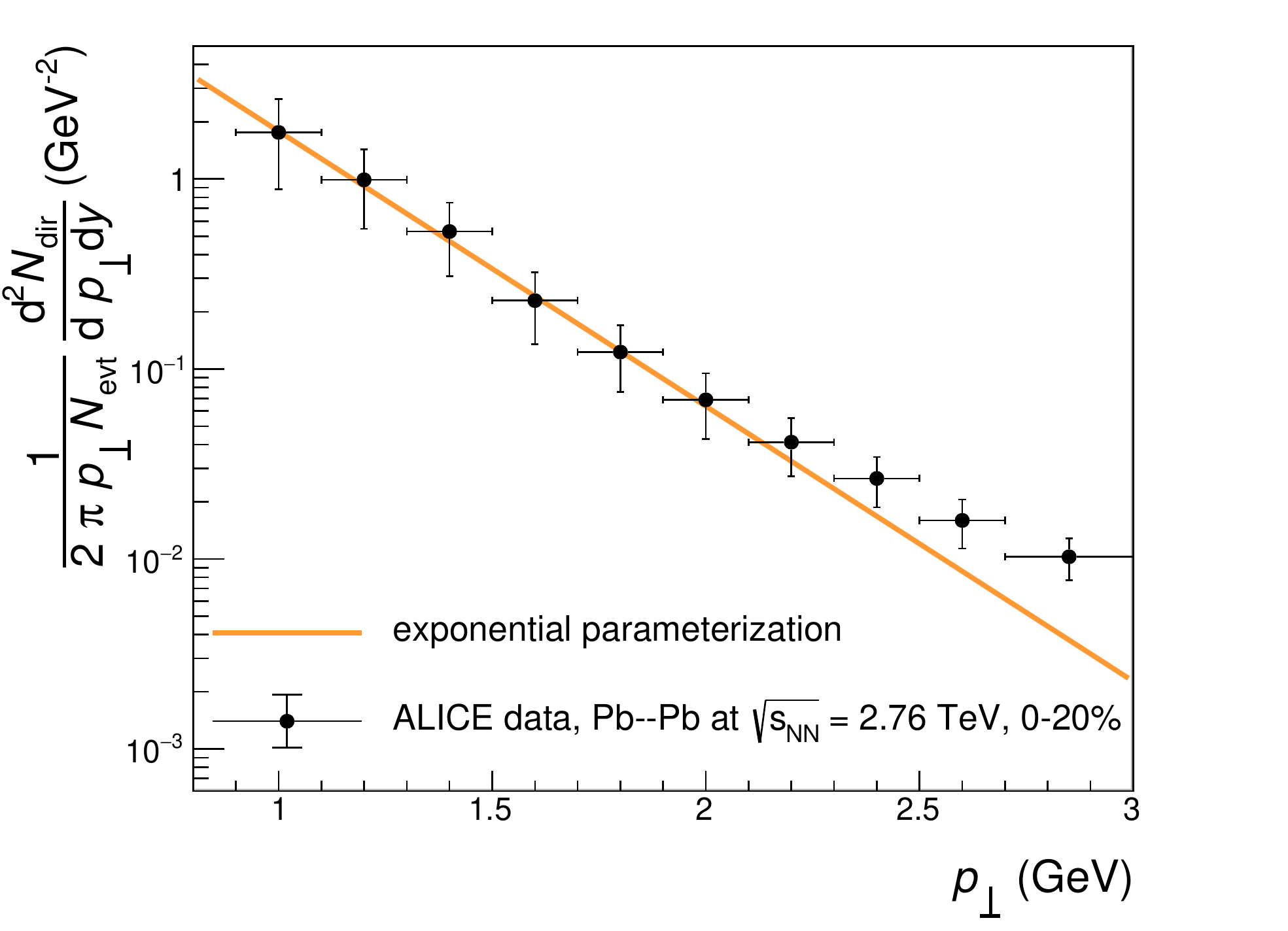}
\caption{\label{fig:ToyFitBackground} Simple exponential parameterization of the direct-photon spectrum in 0--20\% Pb--Pb collisions at $\sqrt{s_\mathrm{NN}} = 2.76\,$TeV \cite{ALICE}. The chosen inverse slope parameter is $T = 0.3\,$GeV.}
\end{figure}

The basis for our estimate is the direct-photon spectrum in 0--20\% Pb--Pb collisions at $\sqrt{s_\mathrm{NN}}=2.76\,\mathrm{TeV}$ measured by ALICE \cite{ALICE}. We parameterize the spectrum by 
\begin{equation}
\left. \frac{1}{2 \pi p_\perp N_\mathrm{evt}} \frac{d^2 N_\mathrm{dir}}{d p_\perp \mathrm{d}y} \right|_{y=0} 
= A \exp \left(- \frac{p_{\mathrm{T}}}{T} \right)
\label{eq:dir_photon_spectrum_fit}
\end{equation}
where the inverse slope parameter is set to $T=0.3\,$GeV, see Fig.~\ref{fig:ToyFitBackground}.

From this simple parametrization of the measured direct-photon spectrum we calculate the number $N_\mathrm{p,u}^\mathrm{dir}$ of uncorrelated pairs of direct photons per event in a given $\ql$ bin. We consider a measurement of $C(\ql)$ in 10 MeV wide bins for $|\qo| < 30\,$MeV and $|\qs| < 30\,$MeV in various intervals of pair transverse momentum $\KTS$. 

\begin{figure}
\includegraphics[scale=0.65]{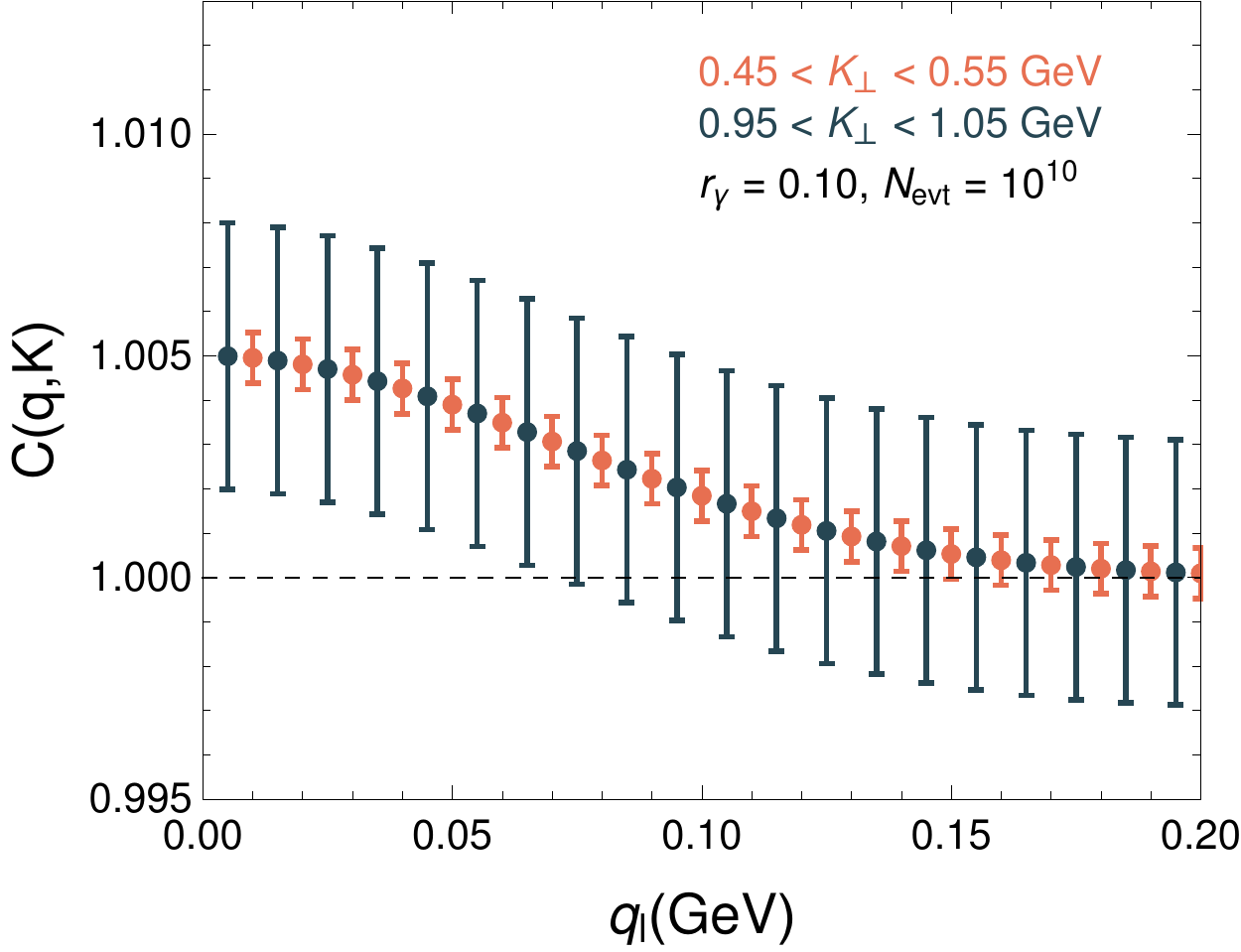}
\caption{\label{fig:HBT_ALICE_Stat} Projected statistical uncertainties for a measurement of $C(\ql)$ in 10 MeV wide bins for two pair transverse momentum ranges $0.45 < \KTS < 0.55\,\mathrm{GeV}$ (in red) and $0.95 < \KTS < 1.05\,\mathrm{GeV}$ (in black) in $10^{10}$ Pb--Pb collisions at $\sqrt{s_\mathrm{NN}} = 2.76\,\mathrm{TeV}$ in the centrality range 0--20\%. The other components of the momentum difference are constrained to $|\qo| < 30\,$MeV, $|\qs| < 30\,\mathrm{MeV}$. The shown correlation function corresponds to a Gaussian parameterization (Eq.~\ref{eq:corr_func_gauss}) with an arbitrarily chosen radius $\Rl = 2\,\mathrm{fm}$.}
\end{figure}

The statistical uncertainty of the total number $C \cdot N_\mathrm{evt}  \cdot N_\mathrm{p,u}^\mathrm{inc}$ of pairs of inclusive photons should be much smaller than the number of pairs $(C-1) \cdot N_\mathrm{evt}  \cdot N_\mathrm{p,u}^\mathrm{inc}$ above the uncorrelated background. Here $N_\mathrm{evt}$ denotes the number of considered Pb--Pb collisions. This corresponds to
\begin{equation}
\sqrt{C \cdot N_\mathrm{evt} \cdot N_\mathrm{p,u}^\mathrm{inc}} \ll 
N_\mathrm{evt} \cdot(C-1) \cdot N_\mathrm{p,u}^\mathrm{inc}.
\end{equation}
Neglecting the small $\sqrt{C}$ term on the left hand-side, the criterion for a significant measurement in the considered bin reads
\begin{equation}
\sigma_\mathrm{rel}^\mathrm{inc} \ll C - 1 = \frac{1}{2} r_\gamma^2 
\end{equation}
where 
\begin{equation}
\sigma_\mathrm{rel}^\mathrm{inc} = 
\frac{1}{\sqrt{N_\mathrm{evt} N_\mathrm{p,u}^\mathrm{inc}}} =
\frac{r_\gamma}{\sqrt{N_\mathrm{evt} N_\mathrm{p,u}^\mathrm{dir}}}.
\end{equation}
Results for the statistical uncertainty $\sigma_\mathrm{rel}^\mathrm{inc}$ of the measured correlation $C(\ql)$ for inclusive photons for $N_\mathrm{evt} = 10^{10}$ are given in Table~\ref{tab:Stat}. This table also shows the ratio $s = 2 \sigma_\mathrm{rel}^\mathrm{inc} / r_\gamma^2$. A value $s \ll 1$ corresponds to a significant measurement. We consider the case of a full photon detection efficiency (1) and the case of a limited detection efficiency (2).
\begin{table}
  \begin{tabular}{c c c c c}
  $\KTS$ (GeV) & $\sigma_\mathrm{rel,1}^\mathrm{inc}$ (\%) & $s_1$ & $\sigma_\mathrm{rel,2}^\mathrm{inc}$ (\%) &$s_2$  \\
  \hline
 \text{0.15--0.25} & 0.001 & 0.002 & 0.021 & 0.043 \\
 \text{0.45--0.55} & 0.002 & 0.005 & 0.057 & 0.114 \\
 \text{0.95--1.05} & 0.012 & 0.024 & 0.299 & 0.600 \\
 \text{1.45--1.55} & 0.063 & 0.127 & 1.580 & 3.170 \\
  \end{tabular}
  \caption{Projected relative statistical uncertainties for $C(\ql)$ measured for pairs of inclusive photons in a 10 MeV wide $\ql$ bin in $10^{10}$ Pb--Pb collisions (centrality 0--20\% ) at 2.76~TeV in one unit around midrapidity ($|y| < 0.5$). The other two components of the pair momentum difference are constrained to $|\qo| < 30\,$MeV and $|\qs| < 30\,$MeV. The uncertainty $\sigma_\mathrm{rel,1}^\mathrm{inc}$ corresponds to a 100\% photon detection efficiency. For $\sigma_\mathrm{rel,2}^\mathrm{inc}$ a photon detection efficiency of $\varepsilon = p_\mathrm{conv} \times \varepsilon_\mathrm{reco} = 0.04$ is assumed where $p_\mathrm{conv} = 0.08$ and $\varepsilon_\mathrm{reco} = 0.5$ roughly correspond to the photon conversion and reconstruction efficiencies in the photon conversion measurements of the ALICE experiment \cite{ALICE}. The table also shows the ratio $s = 2 \sigma_\mathrm{rel}^\mathrm{inc} / r_\gamma^2$ for these two cases. For a significant measurement $s$ needs to be significantly smaller than unity.}
  \label{tab:Stat}
\end{table}
From Table~\ref{tab:Stat} one can conclude that with $N_\mathrm{evt} = 10^{10}$ Pb--Pb collisions there is enough statistics to measure direct-photon HBT correlations up to a pair transverse momentum of $\KTS \approx 1\,\mathrm{GeV}$. For this value of $\KTS$ we illustrate the projected statistical uncertainties of $C$ measured for pairs of inclusive photons in black in Fig.~\ref{fig:HBT_ALICE_Stat}. For comparison the distribution is also shown in red for $\KTS \approx 0.5\,\mathrm{GeV}$, which has much smaller projected statistical uncertainties. This provides a motivation to experimentally explore photon HBT correlation in the upcoming high-luminosity LHC runs \cite{Citron:2018lsq} and to study in detail all sources of systematic uncertainties which might affect the measurement.

\section{\label{sec:Summary}Summary and Conclusion}

In this work, we present a case study of photon interferometry 
 exploring the space-time evolution of the fireball to
investigate possible new photon sources.
In addition to standard thermal and prompt
photons, 
we consider two different scenarios, one in which
 additional photons
are produced from the early pre-equilibrium stage, and one in which
the thermal rates are enhanced close to the transition.
In both cases the mid-rapidity direct photon yields agree with ALICE results in
central (0-20\%) Pb-Pb collisions at $\sqrt{s_{\mathrm{NN}}} = 2.76\,\text{TeV}$.

We then
compute the HBT correlators in the diagonal directions, 
$\qo, \,\qs$ and $\ql$ for different  transverse pair momenta. 
In general, including photon emission from the pre-equilibrium stage 
widens the correlation because of a more compact emission source at early times.
Conversely, the late-time enhancement makes the two-photon correlation narrower.
From these correlators we extract the HBT radii    $\Rl$, $\Ro$, and  $\Rs$.
The longitudinal radius exhibits the largest 
difference between the thermal and the other two scenarios, namely $\sim 80\%$ and  $\sim 20\%$ for early and late time enhancements.
In comparison, the $\Ro$ and $\Rs$ radii
are only mildly affected, with $\sim 20\%$ and $\sim 5\%$ changes respectively.

Direct photons see the entire space-time evolution of the expanding
fireball, which results in pronounced 
non\nobreakdash-Gaussianities in the photon
 HBT signal.
 To quantify these effects, we compute the normalized excess kurtosis, which we find to be 
 largest for the longitudinal direction and significantly smaller in 
 the outward and sideward directions. At small transverse momentum, the significant
 differences in the observed longitudinal non-Gaussianities 
  provide a striking new signature sensitive to
 the different photon emission sources.

In view of the potential of two-photon correlation measurements, we perform an experimental
feasibility study.
With the projected count of
$\sim 10^{10}$ heavy-ion events at the upcoming LHC Runs 3 and 4, we determine the statistical uncertainties
of the experimental signal.
Owing 
to the photons from neutral meson decays,
the HBT signal is attenuated to a percent level. For 
transverse momenta $\KTS \lesssim0.5\, \text{GeV}$
statistics will be sufficient for the measurement
of the correlation function. 
However, the differences between the early and late
time scenarios are most pronounced at higher photon-pair momenta, where statistical uncertainties
are large. Therefore, it is unlikely that
the photon interferometry alone can be used to identify the correct photon emission scenario. 
Nevertheless, we show that photon HBT signal is an experimentally accessible observable with
sensitivity to photon production physics. In conjunction with other
observables, e.g.\ elliptic flow, HBT correlations could be used to rule out certain
models and, therefore, 
motivate further theoretical studies and experimental estimates of systematic errors.

\appendix
\section*{Acknowledgements}
Authors thank Eduardo Grossi, Jean-François Paquet, and Johanna Stachel for valuable discussions. 
We thank Prithwish Tribedy for sharing IP-Glasma simulation results and Werner Vogelsang for sharing
the pp prompt photon data.
OGM is funded by HGS-HIRe. This work is part of and supported by the DFG Collaborative Research Centre "SFB 1225 (ISOQUANT)". 

\section{Thermal rates \label{app:rates}} 

After the thermalization of the colored medium, photons can be emitted from either a thermalized quark-gluon-plasma or can be produced by hadronic processes in the hadron resonance gas phase. In the following we will summarize the rates used in this work to compute the radiation from the thermal phases. 

\subsection*{Thermal rate for Quark Gluon Plasma}

As indicated above, to emit photons from the thermal QGP phase we will use the full LO rate of Ref.~\cite{Arnold:2001ms}, which was computed using weak-coupling expansion in a thermal QFT. The rate used is,    

\begin{equation}
E\frac{d\,N }{d^4\,X\,d^3\,p }=A(p)\,\nu_{LO}\left(\frac{p}{T}\right)
\end{equation}

with the leading-log coefficient ${A}(p)$, which is given by 
\begin{equation}
A(p)= \frac{2\,\alpha}{(2\pi)^{3}} d_F \left[ \sum_c q_c^2\right]\,m^2_D\,  f_{q,eq}\left(\frac{E}{T}\right)
\end{equation}

The remaining part of this rate is given by 
\begin{equation}
\begin{split}
\nu_{LO}\left(x\right) \equiv & \frac{1}{2} \ln\left(2x\right) + C_{2 \leftrightarrow 2} \left(x\right) \\
&+ C_{\mathrm{brem}} \left(x\right)  + C_{\mathrm{annih}} \left(x\right)
\end{split}
\label{eq:L_LO}
\end{equation}
\noindent
with the Fermi distribution function $n_{f}(k)=\left[\exp(k/T)+1\right]^{-1} $. The dimension of the quark representation is $d_{\mathrm{F}} $, which is 3 in our case. Summing over the charges of quarks,  $q_{s}$, one gets  $d_{\mathrm{F}} \sum_{s} q_{s}^{2} = 3 \times (1 \cdot (2/3)^2 + 2 \cdot (1/3)^2 ) = 3 \times 6/9$. The leading-order asymptotic thermal quark-mass $m_{\infty}$ is given by~\cite{Weldon:1982bn} to be

\begin{equation}
m^{2}_{\infty} = \frac{C_{\mathrm{F}} g_{s}^2T^2}{4}
\end{equation}

\noindent
with the quadratic Casimir of the quark representation $C_{\mathrm{F}}$, which is $C_{\mathrm{F}}=4/3$ for QCD, and the strong coupling $g_{s} = \sqrt{4\pi\,\alpha_{s}}$. Using the running coupling prescription, 
\begin{equation}
\alpha_s(Q) = \frac{12\pi}{(11N_c -2N_f)\log(Q^2/\Lambda^2_{QCD})}
\end{equation}
where the cutoff scale, $\Lambda_{QCD}=0.2\GeV$. For $SU(3)$, with $N_c=3$ and three flavours, $N_f=3$ we get that for ALICE energies, $\alpha_{s}\approx 0.3$. The functions that describe the two-to-two particle processes ($C_{2 \leftrightarrow 2} $) and the in-medium bremsstrahlung ($C_{\mathrm{brem}}$) and annihilation ($C_{\mathrm{annih}}$) processes are,
\small
\begin{equation}
\begin{split}
C_{2 \leftrightarrow 2} &= 0.041 x^{-1} - 0.3615 + 1.01e^{-1.35x}\\
C_{\mathrm{brem}} &+ C_{\mathrm{annih}} \simeq \sqrt{1 + \frac{1}{6} \mathrm{N_{f}}}  \\ 
&\times \left[\frac{0.548 \log(12.28+1/x)}{x^{2/3}} + \frac{0.133x}{\sqrt{1+x/16.27}} \right]
\end{split}
\end{equation} 
\normalsize
with $x=p/T$ for three flavours, $N_{f}= 3$.
These functions were obtained by approximating the full kinetic kernels. The full logarithm under the log  will also be used to enhance the non-equilibrium rate, with the substitution $x=E/t \rightarrow x'=E/Q$. 
\subsection*{\label{sec:HRG}Photon emission from the hadron resonance gas}

For from the hadron resonance gas (HRG) phase, we use the the thermal photon emission rate  the parametrization Ref.~\cite{Heffernan:2014mla}. These  parametrizations have an underlying error of no more than 20\% with the microscopic calculated values \cite{Rapp:1999us,Liu:2007zzw}. We use this parametrization since the inclusion of the full cross section into a phenomenological model is not practical, and very computationally expensive \cite{Heffernan:2014mla}. Two different contributions are included, one from the meson channel $\pi\pi\rightarrow \pi\pi\gamma$ and another one including  the emission from  in-medium $\rho$ mesons. These parametrizations can be applied to photons with energies $q_{0}$ between $0.2$ and $5\GeV$, at temperatures between $T=100-180\MeV$ and baryon chemical potentials of $\mu_B=0-400\MeV$. For these investigations we will set $\mu_B=0$.

The contribution from in-medium $\rho$-mesons, including channels like $\pi N \rightarrow\pi N \gamma$ and $N N \rightarrow N N \gamma$, are universally given by \cite{Heffernan:2014mla},

\begin{equation}
q_{0} \frac{dR_{\gamma}^{\rho}}{d^{3}q}(q_{0};T) = \exp\left[a(T)q_{0} + b(T) + \frac{c(T)}{q_{0}+0.2}\right]\,.
\end{equation}
Here, and in the following, $q_0$ and $T$ are given in units of GeV. We will use the fitted  parameters given in Ref.~\cite{Heffernan:2014mla}
\begin{eqnarray}
&&a(T) = -31.21 + 353.61 T - 1739.4 T^2 + 3105 T^3\nonumber\\ 
&&b(T) = -5.513 -42.2T + 333T^2-570T^3\\
&&c(T)=-6.153+57T-134.61T^2+8.31T^3\nonumber
\end{eqnarray}

Nevertheless, this contribution does not include meson-meson bremsstrahlung, strongly dominated by the $\pi \pi \rightarrow \pi \pi \gamma$ channel. The contribution from  $\pi K$ scattering is subleading, and will not be included, since it comprises  at most an increase of $20\%$. The following fit function is used

\begin{equation}
\begin{split}
q_{0} \frac{dR_{\gamma}^{Brems}}{d^{3}q}(q_{0};T) =& \exp\{\alpha_{B}(T)+ q_{0}\, \beta_{B}(T)\\
 +& \gamma_{B}\,q_{0}^2 + \delta_{B}(T)(q_{0}+0.2)^{-1}\}
 \end{split}
\end{equation}
with the following fitted parameters, 

\small
\begin{equation}
\begin{split}
\alpha_{B}(T) &= -16.28 + 62.45 T - 93.4T^2 + 7.5 T^3\\ 
\beta_{B}(T) &= -35.54 +414.8T -2054T^2+3718.8T^3\\
\gamma_{B}(T)&=0.7364-10.72T+56.32^2-103.5T^3\\
\delta_{B}(T)&=-2.51+58.152T-318.24T^2+610.7T^3
\end{split}
\end{equation}
\normalsize
In the HRG, these two contributions are relevant for different kinematic windows of the photons. For a temperature of $150\MeV$ , soft photons ($q_{0} < 0.4\GeV$) are strongly dominated by $\pi\pi$ scattering. On the other hand, the contribution form $\rho$-meson decays is an order of magnitude larger for $q_{0} > 1\GeV$~\cite{Liu:2007zzw}. 
\bibliography{References}

\begin{thebibliography}{68}%
\makeatletter
\providecommand \@ifxundefined [1]{%
 \@ifx{#1\undefined}
}%
\providecommand \@ifnum [1]{%
 \ifnum #1\expandafter \@firstoftwo
 \else \expandafter \@secondoftwo
 \fi
}%
\providecommand \@ifx [1]{%
 \ifx #1\expandafter \@firstoftwo
 \else \expandafter \@secondoftwo
 \fi
}%
\providecommand \natexlab [1]{#1}%
\providecommand \enquote  [1]{``#1''}%
\providecommand \bibnamefont  [1]{#1}%
\providecommand \bibfnamefont [1]{#1}%
\providecommand \citenamefont [1]{#1}%
\providecommand \href@noop [0]{\@secondoftwo}%
\providecommand \href [0]{\begingroup \@sanitize@url \@href}%
\providecommand \@href[1]{\@@startlink{#1}\@@href}%
\providecommand \@@href[1]{\endgroup#1\@@endlink}%
\providecommand \@sanitize@url [0]{\catcode `\\12\catcode `\$12\catcode
  `\&12\catcode `\#12\catcode `\^12\catcode `\_12\catcode `\%12\relax}%
\providecommand \@@startlink[1]{}%
\providecommand \@@endlink[0]{}%
\providecommand \url  [0]{\begingroup\@sanitize@url \@url }%
\providecommand \@url [1]{\endgroup\@href {#1}{\urlprefix }}%
\providecommand \urlprefix  [0]{URL }%
\providecommand \Eprint [0]{\href }%
\providecommand \doibase [0]{http://dx.doi.org/}%
\providecommand \selectlanguage [0]{\@gobble}%
\providecommand \bibinfo  [0]{\@secondoftwo}%
\providecommand \bibfield  [0]{\@secondoftwo}%
\providecommand \translation [1]{[#1]}%
\providecommand \BibitemOpen [0]{}%
\providecommand \bibitemStop [0]{}%
\providecommand \bibitemNoStop [0]{.\EOS\space}%
\providecommand \EOS [0]{\spacefactor3000\relax}%
\providecommand \BibitemShut  [1]{\csname bibitem#1\endcsname}%
\let\auto@bib@innerbib\@empty
\bibitem [{\citenamefont {Busza}\ \emph {et~al.}(2018)\citenamefont {Busza},
  \citenamefont {Rajagopal},\ and\ \citenamefont {van~der
  Schee}}]{Busza:2018rrf}%
  \BibitemOpen
  \bibfield  {author} {\bibinfo {author} {\bibfnamefont {W.}~\bibnamefont
  {Busza}}, \bibinfo {author} {\bibfnamefont {K.}~\bibnamefont {Rajagopal}}, \
  and\ \bibinfo {author} {\bibfnamefont {W.}~\bibnamefont {van~der Schee}},\
  }\href {\doibase 10.1146/annurev-nucl-101917-020852} {\bibfield  {journal}
  {\bibinfo  {journal} {Ann. Rev. Nucl. Part. Sci.}\ }\textbf {\bibinfo
  {volume} {68}},\ \bibinfo {pages} {339} (\bibinfo {year} {2018})},\ \Eprint
  {http://arxiv.org/abs/1802.04801} {arXiv:1802.04801 [hep-ph]} \BibitemShut
  {NoStop}%
\bibitem [{\citenamefont {Chatterjee}\ \emph {et~al.}(2010)\citenamefont
  {Chatterjee}, \citenamefont {Bhattacharya},\ and\ \citenamefont
  {Srivastava}}]{Chatterjee:2009rs}%
  \BibitemOpen
  \bibfield  {author} {\bibinfo {author} {\bibfnamefont {R.}~\bibnamefont
  {Chatterjee}}, \bibinfo {author} {\bibfnamefont {L.}~\bibnamefont
  {Bhattacharya}}, \ and\ \bibinfo {author} {\bibfnamefont {D.~K.}\
  \bibnamefont {Srivastava}},\ }\bibfield  {booktitle} {\emph {\bibinfo
  {booktitle} {{QGP Winter School 2008 Jaipur, India, February 1-3, 2008}}},\
  }\href {\doibase 10.1007/978-3-642-02286-9_7} {\bibfield  {journal} {\bibinfo
   {journal} {Lect. Notes Phys.}\ }\textbf {\bibinfo {volume} {785}},\ \bibinfo
  {pages} {219} (\bibinfo {year} {2010})},\ \Eprint
  {http://arxiv.org/abs/0901.3610} {arXiv:0901.3610 [nucl-th]} \BibitemShut
  {NoStop}%
\bibitem [{\citenamefont {Adare}\ \emph {et~al.}(2015)\citenamefont {Adare}
  \emph {et~al.}}]{Adare:2014fwh}%
  \BibitemOpen
  \bibfield  {author} {\bibinfo {author} {\bibfnamefont {A.}~\bibnamefont
  {Adare}} \emph {et~al.} (\bibinfo {collaboration} {PHENIX}),\ }\href
  {\doibase 10.1103/PhysRevC.91.064904} {\bibfield  {journal} {\bibinfo
  {journal} {Phys. Rev.}\ }\textbf {\bibinfo {volume} {C91}},\ \bibinfo {pages}
  {064904} (\bibinfo {year} {2015})},\ \Eprint {http://arxiv.org/abs/1405.3940}
  {arXiv:1405.3940 [nucl-ex]} \BibitemShut {NoStop}%
\bibitem [{\citenamefont {Adare}\ \emph {et~al.}(2016)\citenamefont {Adare}
  \emph {et~al.}}]{Adare:2015lcd}%
  \BibitemOpen
  \bibfield  {author} {\bibinfo {author} {\bibfnamefont {A.}~\bibnamefont
  {Adare}} \emph {et~al.} (\bibinfo {collaboration} {PHENIX}),\ }\href
  {\doibase 10.1103/PhysRevC.94.064901} {\bibfield  {journal} {\bibinfo
  {journal} {Phys. Rev.}\ }\textbf {\bibinfo {volume} {C94}},\ \bibinfo {pages}
  {064901} (\bibinfo {year} {2016})},\ \Eprint
  {http://arxiv.org/abs/1509.07758} {arXiv:1509.07758 [nucl-ex]} \BibitemShut
  {NoStop}%
\bibitem [{\citenamefont {Paquet}\ \emph {et~al.}(2016)\citenamefont {Paquet},
  \citenamefont {Shen}, \citenamefont {Denicol}, \citenamefont {Luzum},
  \citenamefont {Schenke}, \citenamefont {Jeon},\ and\ \citenamefont
  {Gale}}]{Paquet:2015lta}%
  \BibitemOpen
  \bibfield  {author} {\bibinfo {author} {\bibfnamefont {J.-F.}\ \bibnamefont
  {Paquet}}, \bibinfo {author} {\bibfnamefont {C.}~\bibnamefont {Shen}},
  \bibinfo {author} {\bibfnamefont {G.~S.}\ \bibnamefont {Denicol}}, \bibinfo
  {author} {\bibfnamefont {M.}~\bibnamefont {Luzum}}, \bibinfo {author}
  {\bibfnamefont {B.}~\bibnamefont {Schenke}}, \bibinfo {author} {\bibfnamefont
  {S.}~\bibnamefont {Jeon}}, \ and\ \bibinfo {author} {\bibfnamefont
  {C.}~\bibnamefont {Gale}},\ }\href {\doibase 10.1103/PhysRevC.93.044906}
  {\bibfield  {journal} {\bibinfo  {journal} {Phys. Rev.}\ }\textbf {\bibinfo
  {volume} {C93}},\ \bibinfo {pages} {044906} (\bibinfo {year} {2016})},\
  \Eprint {http://arxiv.org/abs/1509.06738} {arXiv:1509.06738 [hep-ph]}
  \BibitemShut {NoStop}%
\bibitem [{\citenamefont {Adam}\ \emph {et~al.}(2016)\citenamefont {Adam} \emph
  {et~al.}}]{ALICE}%
  \BibitemOpen
  \bibfield  {author} {\bibinfo {author} {\bibfnamefont {J.}~\bibnamefont
  {Adam}} \emph {et~al.},\ }\href {\doibase
  https://doi.org/10.1016/j.physletb.2016.01.020} {\bibfield  {journal}
  {\bibinfo  {journal} {Physics Letters B}\ }\textbf {\bibinfo {volume}
  {754}},\ \bibinfo {pages} {235 } (\bibinfo {year} {2016})}\BibitemShut
  {NoStop}%
\bibitem [{\citenamefont {Acharya}\ \emph {et~al.}(2019)\citenamefont {Acharya}
  \emph {et~al.}}]{Acharya:2018bdy}%
  \BibitemOpen
  \bibfield  {author} {\bibinfo {author} {\bibfnamefont {S.}~\bibnamefont
  {Acharya}} \emph {et~al.} (\bibinfo {collaboration} {ALICE}),\ }\href
  {\doibase 10.1016/j.physletb.2018.11.039} {\bibfield  {journal} {\bibinfo
  {journal} {Phys. Lett.}\ }\textbf {\bibinfo {volume} {B789}},\ \bibinfo
  {pages} {308} (\bibinfo {year} {2019})},\ \Eprint
  {http://arxiv.org/abs/1805.04403} {arXiv:1805.04403 [nucl-ex]} \BibitemShut
  {NoStop}%
\bibitem [{\citenamefont {David}(2019)}]{David:2019wpt}%
  \BibitemOpen
  \bibfield  {author} {\bibinfo {author} {\bibfnamefont {G.}~\bibnamefont
  {David}},\ }\href@noop {} {\  (\bibinfo {year} {2019})},\ \Eprint
  {http://arxiv.org/abs/1907.08893} {arXiv:1907.08893 [nucl-ex]} \BibitemShut
  {NoStop}%
\bibitem [{\citenamefont {Heinz}(1996)}]{Heinz:1996bs}%
  \BibitemOpen
  \bibfield  {author} {\bibinfo {author} {\bibfnamefont {U.~W.}\ \bibnamefont
  {Heinz}},\ }in\ \href@noop {} {\emph {\bibinfo {booktitle} {{Correlations and
  clustering phenomena in subatomic physics. Proceedings, NATO Advanced Study
  Institute, Dronten, Netherlands, August 5-16, 1996}}}}\ (\bibinfo {year}
  {1996})\ pp.\ \bibinfo {pages} {137--177},\ \Eprint
  {http://arxiv.org/abs/nucl-th/9609029} {arXiv:nucl-th/9609029 [nucl-th]}
  \BibitemShut {NoStop}%
\bibitem [{\citenamefont {Hanbury~Brown}\ and\ \citenamefont
  {Twiss}(1954)}]{HanburyBrown:1954amm}%
  \BibitemOpen
  \bibfield  {author} {\bibinfo {author} {\bibfnamefont {R.}~\bibnamefont
  {Hanbury~Brown}}\ and\ \bibinfo {author} {\bibfnamefont {R.~Q.}\ \bibnamefont
  {Twiss}},\ }\href {\doibase 10.1080/14786440708520475} {\bibfield  {journal}
  {\bibinfo  {journal} {Phil. Mag. Ser.7}\ }\textbf {\bibinfo {volume} {45}},\
  \bibinfo {pages} {663} (\bibinfo {year} {1954})}\BibitemShut {NoStop}%
\bibitem [{\citenamefont {Hanbury~Brown}\ and\ \citenamefont
  {Twiss}(1956)}]{HanburyBrown:1956bqd}%
  \BibitemOpen
  \bibfield  {author} {\bibinfo {author} {\bibfnamefont {R.}~\bibnamefont
  {Hanbury~Brown}}\ and\ \bibinfo {author} {\bibfnamefont {R.~Q.}\ \bibnamefont
  {Twiss}},\ }\href {\doibase 10.1038/1781046a0} {\bibfield  {journal}
  {\bibinfo  {journal} {Nature}\ }\textbf {\bibinfo {volume} {178}},\ \bibinfo
  {pages} {1046} (\bibinfo {year} {1956})}\BibitemShut {NoStop}%
\bibitem [{\citenamefont {F{\"o}lling}\ \emph {et~al.}(2005)\citenamefont
  {F{\"o}lling}, \citenamefont {Gerbier}, \citenamefont {Widera}, \citenamefont
  {Mandel}, \citenamefont {Gericke},\ and\ \citenamefont {Bloch}}]{CAHBT1}%
  \BibitemOpen
  \bibfield  {author} {\bibinfo {author} {\bibfnamefont {S.}~\bibnamefont
  {F{\"o}lling}}, \bibinfo {author} {\bibfnamefont {F.}~\bibnamefont
  {Gerbier}}, \bibinfo {author} {\bibfnamefont {A.}~\bibnamefont {Widera}},
  \bibinfo {author} {\bibfnamefont {O.}~\bibnamefont {Mandel}}, \bibinfo
  {author} {\bibfnamefont {T.}~\bibnamefont {Gericke}}, \ and\ \bibinfo
  {author} {\bibfnamefont {I.}~\bibnamefont {Bloch}},\ }\href {\doibase
  10.1038/nature03500} {\bibfield  {journal} {\bibinfo  {journal} {Nature}\
  }\textbf {\bibinfo {volume} {434}},\ \bibinfo {pages} {481} (\bibinfo {year}
  {2005})}\BibitemShut {NoStop}%
\bibitem [{\citenamefont {Rom}\ \emph {et~al.}(2006)\citenamefont {Rom},
  \citenamefont {Best}, \citenamefont {van Oosten}, \citenamefont {Schneider},
  \citenamefont {F{\"o}lling}, \citenamefont {Paredes},\ and\ \citenamefont
  {Bloch}}]{CAHBT2}%
  \BibitemOpen
  \bibfield  {author} {\bibinfo {author} {\bibfnamefont {T.}~\bibnamefont
  {Rom}}, \bibinfo {author} {\bibfnamefont {T.}~\bibnamefont {Best}}, \bibinfo
  {author} {\bibfnamefont {D.}~\bibnamefont {van Oosten}}, \bibinfo {author}
  {\bibfnamefont {U.}~\bibnamefont {Schneider}}, \bibinfo {author}
  {\bibfnamefont {S.}~\bibnamefont {F{\"o}lling}}, \bibinfo {author}
  {\bibfnamefont {B.}~\bibnamefont {Paredes}}, \ and\ \bibinfo {author}
  {\bibfnamefont {I.}~\bibnamefont {Bloch}},\ }\href {\doibase
  10.1038/nature05319} {\bibfield  {journal} {\bibinfo  {journal} {Nature}\
  }\textbf {\bibinfo {volume} {444}},\ \bibinfo {pages} {733} (\bibinfo {year}
  {2006})}\BibitemShut {NoStop}%
\bibitem [{\citenamefont {Gyulassy}\ \emph {et~al.}(1979)\citenamefont
  {Gyulassy}, \citenamefont {Kauffmann},\ and\ \citenamefont
  {Wilson}}]{Gyulassy:1979yi}%
  \BibitemOpen
  \bibfield  {author} {\bibinfo {author} {\bibfnamefont {M.}~\bibnamefont
  {Gyulassy}}, \bibinfo {author} {\bibfnamefont {S.~K.}\ \bibnamefont
  {Kauffmann}}, \ and\ \bibinfo {author} {\bibfnamefont {L.~W.}\ \bibnamefont
  {Wilson}},\ }\href {\doibase 10.1103/PhysRevC.20.2267} {\bibfield  {journal}
  {\bibinfo  {journal} {Phys. Rev.}\ }\textbf {\bibinfo {volume} {C20}},\
  \bibinfo {pages} {2267} (\bibinfo {year} {1979})}\BibitemShut {NoStop}%
\bibitem [{\citenamefont {Calligarich}\ \emph {et~al.}(1976)\citenamefont
  {Calligarich} \emph {et~al.}}]{Calligarich:1976kp}%
  \BibitemOpen
  \bibfield  {author} {\bibinfo {author} {\bibfnamefont {E.}~\bibnamefont
  {Calligarich}} \emph {et~al.},\ }\href {\doibase 10.1007/BF02746938}
  {\bibfield  {journal} {\bibinfo  {journal} {Lett. Nuovo Cim.}\ }\textbf
  {\bibinfo {volume} {16}},\ \bibinfo {pages} {129} (\bibinfo {year}
  {1976})}\BibitemShut {NoStop}%
\bibitem [{\citenamefont {Pratt}(1984)}]{Pratt:1984su}%
  \BibitemOpen
  \bibfield  {author} {\bibinfo {author} {\bibfnamefont {S.}~\bibnamefont
  {Pratt}},\ }\href {\doibase 10.1103/PhysRevLett.53.1219} {\bibfield
  {journal} {\bibinfo  {journal} {Phys. Rev. Lett.}\ }\textbf {\bibinfo
  {volume} {53}},\ \bibinfo {pages} {1219} (\bibinfo {year}
  {1984})}\BibitemShut {NoStop}%
\bibitem [{\citenamefont {Srivastava}\ and\ \citenamefont
  {Kapusta}(1993)}]{Srivastava:1993js}%
  \BibitemOpen
  \bibfield  {author} {\bibinfo {author} {\bibfnamefont {D.~K.}\ \bibnamefont
  {Srivastava}}\ and\ \bibinfo {author} {\bibfnamefont {J.~I.}\ \bibnamefont
  {Kapusta}},\ }\bibfield  {booktitle} {\emph {\bibinfo {booktitle} {{Borlange
  Quark Matter 1993:0523-526}}},\ }\href {\doibase
  10.1016/0370-2693(93)90183-I} {\bibfield  {journal} {\bibinfo  {journal}
  {Phys. Lett.}\ }\textbf {\bibinfo {volume} {B307}},\ \bibinfo {pages} {1}
  (\bibinfo {year} {1993})}\BibitemShut {NoStop}%
\bibitem [{\citenamefont {Timmermann}\ \emph {et~al.}(1994)\citenamefont
  {Timmermann}, \citenamefont {Plumer}, \citenamefont {Razumov},\ and\
  \citenamefont {Weiner}}]{Timmermann:1994kb}%
  \BibitemOpen
  \bibfield  {author} {\bibinfo {author} {\bibfnamefont {A.}~\bibnamefont
  {Timmermann}}, \bibinfo {author} {\bibfnamefont {M.}~\bibnamefont {Plumer}},
  \bibinfo {author} {\bibfnamefont {L.}~\bibnamefont {Razumov}}, \ and\
  \bibinfo {author} {\bibfnamefont {R.~M.}\ \bibnamefont {Weiner}},\ }\href
  {\doibase 10.1103/PhysRevC.50.3060} {\bibfield  {journal} {\bibinfo
  {journal} {Phys. Rev.}\ }\textbf {\bibinfo {volume} {C50}},\ \bibinfo {pages}
  {3060} (\bibinfo {year} {1994})},\ \Eprint
  {http://arxiv.org/abs/hep-ph/9405232} {arXiv:hep-ph/9405232 [hep-ph]}
  \BibitemShut {NoStop}%
\bibitem [{\citenamefont {Slotta}\ and\ \citenamefont
  {Heinz}(1997)}]{Slotta:1996cf}%
  \BibitemOpen
  \bibfield  {author} {\bibinfo {author} {\bibfnamefont {C.}~\bibnamefont
  {Slotta}}\ and\ \bibinfo {author} {\bibfnamefont {U.~W.}\ \bibnamefont
  {Heinz}},\ }\href {\doibase 10.1016/S0370-2693(96)01504-3} {\bibfield
  {journal} {\bibinfo  {journal} {Phys. Lett.}\ }\textbf {\bibinfo {volume}
  {B391}},\ \bibinfo {pages} {469} (\bibinfo {year} {1997})},\ \Eprint
  {http://arxiv.org/abs/nucl-th/9610016} {arXiv:nucl-th/9610016 [nucl-th]}
  \BibitemShut {NoStop}%
\bibitem [{\citenamefont {Srivastava}(2005)}]{Srivastava:2004xp}%
  \BibitemOpen
  \bibfield  {author} {\bibinfo {author} {\bibfnamefont {D.~K.}\ \bibnamefont
  {Srivastava}},\ }\href {\doibase 10.1103/PhysRevC.71.034905} {\bibfield
  {journal} {\bibinfo  {journal} {Phys. Rev.}\ }\textbf {\bibinfo {volume}
  {C71}},\ \bibinfo {pages} {034905} (\bibinfo {year} {2005})},\ \Eprint
  {http://arxiv.org/abs/nucl-th/0411041} {arXiv:nucl-th/0411041 [nucl-th]}
  \BibitemShut {NoStop}%
\bibitem [{\citenamefont {Bass}\ \emph {et~al.}(2004)\citenamefont {Bass},
  \citenamefont {Muller},\ and\ \citenamefont {Srivastava}}]{Bass:2004de}%
  \BibitemOpen
  \bibfield  {author} {\bibinfo {author} {\bibfnamefont {S.~A.}\ \bibnamefont
  {Bass}}, \bibinfo {author} {\bibfnamefont {B.}~\bibnamefont {Muller}}, \ and\
  \bibinfo {author} {\bibfnamefont {D.~K.}\ \bibnamefont {Srivastava}},\ }\href
  {\doibase 10.1103/PhysRevLett.93.162301} {\bibfield  {journal} {\bibinfo
  {journal} {Phys. Rev. Lett.}\ }\textbf {\bibinfo {volume} {93}},\ \bibinfo
  {pages} {162301} (\bibinfo {year} {2004})},\ \Eprint
  {http://arxiv.org/abs/nucl-th/0404050} {arXiv:nucl-th/0404050 [nucl-th]}
  \BibitemShut {NoStop}%
\bibitem [{\citenamefont {Peressounko}(2003)}]{Peressounko:2003cf}%
  \BibitemOpen
  \bibfield  {author} {\bibinfo {author} {\bibfnamefont {D.}~\bibnamefont
  {Peressounko}},\ }\href {\doibase 10.1103/PhysRevC.67.014905} {\bibfield
  {journal} {\bibinfo  {journal} {Phys. Rev.}\ }\textbf {\bibinfo {volume}
  {C67}},\ \bibinfo {pages} {014905} (\bibinfo {year} {2003})}\BibitemShut
  {NoStop}%
\bibitem [{\citenamefont {Frodermann}\ and\ \citenamefont
  {Heinz}(2009)}]{Frodermann:2009nx}%
  \BibitemOpen
  \bibfield  {author} {\bibinfo {author} {\bibfnamefont {E.}~\bibnamefont
  {Frodermann}}\ and\ \bibinfo {author} {\bibfnamefont {U.}~\bibnamefont
  {Heinz}},\ }\href {\doibase 10.1103/PhysRevC.80.044903} {\bibfield  {journal}
  {\bibinfo  {journal} {Phys. Rev.}\ }\textbf {\bibinfo {volume} {C80}},\
  \bibinfo {pages} {044903} (\bibinfo {year} {2009})},\ \Eprint
  {http://arxiv.org/abs/0907.1292} {arXiv:0907.1292 [nucl-th]} \BibitemShut
  {NoStop}%
\bibitem [{\citenamefont {Ipp}\ and\ \citenamefont
  {Somkuti}(2012)}]{Ipp:2012zb}%
  \BibitemOpen
  \bibfield  {author} {\bibinfo {author} {\bibfnamefont {A.}~\bibnamefont
  {Ipp}}\ and\ \bibinfo {author} {\bibfnamefont {P.}~\bibnamefont {Somkuti}},\
  }\href {\doibase 10.1103/PhysRevLett.109.192301} {\bibfield  {journal}
  {\bibinfo  {journal} {Phys. Rev. Lett.}\ }\textbf {\bibinfo {volume} {109}},\
  \bibinfo {pages} {192301} (\bibinfo {year} {2012})},\ \Eprint
  {http://arxiv.org/abs/1207.0197} {arXiv:1207.0197 [hep-ph]} \BibitemShut
  {NoStop}%
\bibitem [{\citenamefont {Aggarwal}\ \emph {et~al.}(2004)\citenamefont
  {Aggarwal} \emph {et~al.}}]{Aggarwal:2003zy}%
  \BibitemOpen
  \bibfield  {author} {\bibinfo {author} {\bibfnamefont {M.~M.}\ \bibnamefont
  {Aggarwal}} \emph {et~al.} (\bibinfo {collaboration} {WA98}),\ }\href
  {\doibase 10.1103/PhysRevLett.93.022301} {\bibfield  {journal} {\bibinfo
  {journal} {Phys. Rev. Lett.}\ }\textbf {\bibinfo {volume} {93}},\ \bibinfo
  {pages} {022301} (\bibinfo {year} {2004})},\ \Eprint
  {http://arxiv.org/abs/nucl-ex/0310022} {arXiv:nucl-ex/0310022 [nucl-ex]}
  \BibitemShut {NoStop}%
\bibitem [{\citenamefont {Citron}\ \emph {et~al.}(2018)\citenamefont {Citron}
  \emph {et~al.}}]{Citron:2018lsq}%
  \BibitemOpen
  \bibfield  {author} {\bibinfo {author} {\bibfnamefont {Z.}~\bibnamefont
  {Citron}} \emph {et~al.},\ }in\ \href@noop {} {\emph {\bibinfo {booktitle}
  {{HL/HE-LHC Workshop: Workshop on the Physics of HL-LHC, and Perspectives at
  HE-LHC Geneva, Switzerland, June 18-20, 2018}}}}\ (\bibinfo {year} {2018})\
  \Eprint {http://arxiv.org/abs/1812.06772} {arXiv:1812.06772 [hep-ph]}
  \BibitemShut {NoStop}%
\bibitem [{\citenamefont {Berges}\ \emph {et~al.}(2017)\citenamefont {Berges},
  \citenamefont {Reygers}, \citenamefont {Tanji},\ and\ \citenamefont
  {Venugopalan}}]{Berges:2017eom}%
  \BibitemOpen
  \bibfield  {author} {\bibinfo {author} {\bibfnamefont {J.}~\bibnamefont
  {Berges}}, \bibinfo {author} {\bibfnamefont {K.}~\bibnamefont {Reygers}},
  \bibinfo {author} {\bibfnamefont {N.}~\bibnamefont {Tanji}}, \ and\ \bibinfo
  {author} {\bibfnamefont {R.}~\bibnamefont {Venugopalan}},\ }\href {\doibase
  10.1103/PhysRevC.95.054904} {\bibfield  {journal} {\bibinfo  {journal} {Phys.
  Rev.}\ }\textbf {\bibinfo {volume} {C95}},\ \bibinfo {pages} {054904}
  (\bibinfo {year} {2017})},\ \Eprint {http://arxiv.org/abs/1701.05064}
  {arXiv:1701.05064 [nucl-th]} \BibitemShut {NoStop}%
\bibitem [{\citenamefont {Monnai}(2019)}]{Monnai:2019vup}%
  \BibitemOpen
  \bibfield  {author} {\bibinfo {author} {\bibfnamefont {A.}~\bibnamefont
  {Monnai}},\ }\href@noop {} {\  (\bibinfo {year} {2019})},\ \Eprint
  {http://arxiv.org/abs/1907.09266} {arXiv:1907.09266 [nucl-th]} \BibitemShut
  {NoStop}%
\bibitem [{\citenamefont {van Hees}\ \emph {et~al.}(2015)\citenamefont {van
  Hees}, \citenamefont {He},\ and\ \citenamefont {Rapp}}]{vanHees:2014ida}%
  \BibitemOpen
  \bibfield  {author} {\bibinfo {author} {\bibfnamefont {H.}~\bibnamefont {van
  Hees}}, \bibinfo {author} {\bibfnamefont {M.}~\bibnamefont {He}}, \ and\
  \bibinfo {author} {\bibfnamefont {R.}~\bibnamefont {Rapp}},\ }\href {\doibase
  10.1016/j.nuclphysa.2014.09.009} {\bibfield  {journal} {\bibinfo  {journal}
  {Nucl. Phys.}\ }\textbf {\bibinfo {volume} {A933}},\ \bibinfo {pages} {256}
  (\bibinfo {year} {2015})},\ \Eprint {http://arxiv.org/abs/1404.2846}
  {arXiv:1404.2846 [nucl-th]} \BibitemShut {NoStop}%
\bibitem [{\citenamefont {van Hees}\ \emph {et~al.}(2011)\citenamefont {van
  Hees}, \citenamefont {Gale},\ and\ \citenamefont {Rapp}}]{vanHees:2011vb}%
  \BibitemOpen
  \bibfield  {author} {\bibinfo {author} {\bibfnamefont {H.}~\bibnamefont {van
  Hees}}, \bibinfo {author} {\bibfnamefont {C.}~\bibnamefont {Gale}}, \ and\
  \bibinfo {author} {\bibfnamefont {R.}~\bibnamefont {Rapp}},\ }\href {\doibase
  10.1103/PhysRevC.84.054906} {\bibfield  {journal} {\bibinfo  {journal} {Phys.
  Rev.}\ }\textbf {\bibinfo {volume} {C84}},\ \bibinfo {pages} {054906}
  (\bibinfo {year} {2011})},\ \Eprint {http://arxiv.org/abs/1108.2131}
  {arXiv:1108.2131 [hep-ph]} \BibitemShut {NoStop}%
\bibitem [{\citenamefont {Wiedemann}\ and\ \citenamefont
  {Heinz}(1999)}]{Wiedemann:1999qn}%
  \BibitemOpen
  \bibfield  {author} {\bibinfo {author} {\bibfnamefont {U.~A.}\ \bibnamefont
  {Wiedemann}}\ and\ \bibinfo {author} {\bibfnamefont {U.~W.}\ \bibnamefont
  {Heinz}},\ }\href {\doibase 10.1016/S0370-1573(99)00032-0} {\bibfield
  {journal} {\bibinfo  {journal} {Phys. Rept.}\ }\textbf {\bibinfo {volume}
  {319}},\ \bibinfo {pages} {145} (\bibinfo {year} {1999})},\ \Eprint
  {http://arxiv.org/abs/nucl-th/9901094} {arXiv:nucl-th/9901094 [nucl-th]}
  \BibitemShut {NoStop}%
\bibitem [{\citenamefont {Csorgo}(2006)}]{Csorgo:2005gd}%
  \BibitemOpen
  \bibfield  {author} {\bibinfo {author} {\bibfnamefont {T.}~\bibnamefont
  {Csorgo}},\ }\bibfield  {booktitle} {\emph {\bibinfo {booktitle}
  {{Proceedings, 5th International Conference on Physics and Astrophysics of
  Quark Gluon Plasma (ICPAQGP 2005): Salt Lake City, India, February 8-12,
  2005}}},\ }\href {\doibase 10.1088/1742-6596/50/1/031} {\bibfield  {journal}
  {\bibinfo  {journal} {J. Phys. Conf. Ser.}\ }\textbf {\bibinfo {volume}
  {50}},\ \bibinfo {pages} {259} (\bibinfo {year} {2006})},\ \Eprint
  {http://arxiv.org/abs/nucl-th/0505019} {arXiv:nucl-th/0505019 [nucl-th]}
  \BibitemShut {NoStop}%
\bibitem [{\citenamefont {Lisa}\ \emph {et~al.}(2005)\citenamefont {Lisa},
  \citenamefont {Pratt}, \citenamefont {Soltz},\ and\ \citenamefont
  {Wiedemann}}]{Lisa:2005dd}%
  \BibitemOpen
  \bibfield  {author} {\bibinfo {author} {\bibfnamefont {M.~A.}\ \bibnamefont
  {Lisa}}, \bibinfo {author} {\bibfnamefont {S.}~\bibnamefont {Pratt}},
  \bibinfo {author} {\bibfnamefont {R.}~\bibnamefont {Soltz}}, \ and\ \bibinfo
  {author} {\bibfnamefont {U.}~\bibnamefont {Wiedemann}},\ }\href {\doibase
  10.1146/annurev.nucl.55.090704.151533} {\bibfield  {journal} {\bibinfo
  {journal} {Ann. Rev. Nucl. Part. Sci.}\ }\textbf {\bibinfo {volume} {55}},\
  \bibinfo {pages} {357} (\bibinfo {year} {2005})},\ \Eprint
  {http://arxiv.org/abs/nucl-ex/0505014} {arXiv:nucl-ex/0505014 [nucl-ex]}
  \BibitemShut {NoStop}%
\bibitem [{\citenamefont {Chapman}\ \emph
  {et~al.}(1995{\natexlab{a}})\citenamefont {Chapman}, \citenamefont {Nix},\
  and\ \citenamefont {Heinz}}]{Chapman:1995nz}%
  \BibitemOpen
  \bibfield  {author} {\bibinfo {author} {\bibfnamefont {S.}~\bibnamefont
  {Chapman}}, \bibinfo {author} {\bibfnamefont {J.~R.}\ \bibnamefont {Nix}}, \
  and\ \bibinfo {author} {\bibfnamefont {U.~W.}\ \bibnamefont {Heinz}},\ }\href
  {\doibase 10.1103/PhysRevC.52.2694} {\bibfield  {journal} {\bibinfo
  {journal} {Phys. Rev.}\ }\textbf {\bibinfo {volume} {C52}},\ \bibinfo {pages}
  {2694} (\bibinfo {year} {1995}{\natexlab{a}})},\ \Eprint
  {http://arxiv.org/abs/nucl-th/9505032} {arXiv:nucl-th/9505032 [nucl-th]}
  \BibitemShut {NoStop}%
\bibitem [{\citenamefont {Chapman}\ \emph
  {et~al.}(1995{\natexlab{b}})\citenamefont {Chapman}, \citenamefont {Scotto},\
  and\ \citenamefont {Heinz}}]{Chapman:1994ax}%
  \BibitemOpen
  \bibfield  {author} {\bibinfo {author} {\bibfnamefont {S.}~\bibnamefont
  {Chapman}}, \bibinfo {author} {\bibfnamefont {P.}~\bibnamefont {Scotto}}, \
  and\ \bibinfo {author} {\bibfnamefont {U.~W.}\ \bibnamefont {Heinz}},\
  }\href@noop {} {\bibfield  {journal} {\bibinfo  {journal} {Acta Phys. Hung.}\
  }\textbf {\bibinfo {volume} {A1}},\ \bibinfo {pages} {1} (\bibinfo {year}
  {1995}{\natexlab{b}})},\ \Eprint {http://arxiv.org/abs/hep-ph/9409349}
  {arXiv:hep-ph/9409349 [hep-ph]} \BibitemShut {NoStop}%
\bibitem [{\citenamefont {Heinz}\ \emph {et~al.}(1996)\citenamefont {Heinz},
  \citenamefont {Tomasik}, \citenamefont {Wiedemann},\ and\ \citenamefont
  {Wu}}]{Heinz:1996qu}%
  \BibitemOpen
  \bibfield  {author} {\bibinfo {author} {\bibfnamefont {U.~W.}\ \bibnamefont
  {Heinz}}, \bibinfo {author} {\bibfnamefont {B.}~\bibnamefont {Tomasik}},
  \bibinfo {author} {\bibfnamefont {U.~A.}\ \bibnamefont {Wiedemann}}, \ and\
  \bibinfo {author} {\bibfnamefont {Y.~F.}\ \bibnamefont {Wu}},\ }\href
  {\doibase 10.1016/0370-2693(96)00657-0} {\bibfield  {journal} {\bibinfo
  {journal} {Phys. Lett.}\ }\textbf {\bibinfo {volume} {B382}},\ \bibinfo
  {pages} {181} (\bibinfo {year} {1996})},\ \Eprint
  {http://arxiv.org/abs/nucl-th/9603011} {arXiv:nucl-th/9603011 [nucl-th]}
  \BibitemShut {NoStop}%
\bibitem [{\citenamefont {Chapman}\ \emph
  {et~al.}(1995{\natexlab{c}})\citenamefont {Chapman}, \citenamefont {Scotto},\
  and\ \citenamefont {Heinz}}]{Chapman:1994yv}%
  \BibitemOpen
  \bibfield  {author} {\bibinfo {author} {\bibfnamefont {S.}~\bibnamefont
  {Chapman}}, \bibinfo {author} {\bibfnamefont {P.}~\bibnamefont {Scotto}}, \
  and\ \bibinfo {author} {\bibfnamefont {U.~W.}\ \bibnamefont {Heinz}},\ }\href
  {\doibase 10.1103/PhysRevLett.74.4400} {\bibfield  {journal} {\bibinfo
  {journal} {Phys. Rev. Lett.}\ }\textbf {\bibinfo {volume} {74}},\ \bibinfo
  {pages} {4400} (\bibinfo {year} {1995}{\natexlab{c}})},\ \Eprint
  {http://arxiv.org/abs/hep-ph/9408207} {arXiv:hep-ph/9408207 [hep-ph]}
  \BibitemShut {NoStop}%
\bibitem [{\citenamefont {Shen}\ \emph
  {et~al.}(2016{\natexlab{a}})\citenamefont {Shen}, \citenamefont {Qiu},
  \citenamefont {Song}, \citenamefont {Bernhard}, \citenamefont {Bass},\ and\
  \citenamefont {Heinz}}]{VISHNU}%
  \BibitemOpen
  \bibfield  {author} {\bibinfo {author} {\bibfnamefont {C.}~\bibnamefont
  {Shen}}, \bibinfo {author} {\bibfnamefont {Z.}~\bibnamefont {Qiu}}, \bibinfo
  {author} {\bibfnamefont {H.}~\bibnamefont {Song}}, \bibinfo {author}
  {\bibfnamefont {J.}~\bibnamefont {Bernhard}}, \bibinfo {author}
  {\bibfnamefont {S.}~\bibnamefont {Bass}}, \ and\ \bibinfo {author}
  {\bibfnamefont {U.}~\bibnamefont {Heinz}},\ }\href {\doibase
  https://doi.org/10.1016/j.cpc.2015.08.039} {\bibfield  {journal} {\bibinfo
  {journal} {Computer Physics Communications}\ }\textbf {\bibinfo {volume}
  {199}},\ \bibinfo {pages} {61 } (\bibinfo {year}
  {2016}{\natexlab{a}})}\BibitemShut {NoStop}%
\bibitem [{VIS()}]{VISHNUwebsite}%
  \BibitemOpen
  \href@noop {} {\enquote {\bibinfo {title} {Vishnu code package},}\ }\bibinfo
  {howpublished} {\url{https://u.osu.edu/vishnu/downloads/}},\ \bibinfo {note}
  {accessed: 2018-03-14}\BibitemShut {NoStop}%
\bibitem [{\citenamefont {Shen}\ \emph
  {et~al.}(2016{\natexlab{b}})\citenamefont {Shen}, \citenamefont {Qiu},
  \citenamefont {Song}, \citenamefont {Bernhard}, \citenamefont {Bass},\ and\
  \citenamefont {Heinz}}]{Shen:2014vra}%
  \BibitemOpen
  \bibfield  {author} {\bibinfo {author} {\bibfnamefont {C.}~\bibnamefont
  {Shen}}, \bibinfo {author} {\bibfnamefont {Z.}~\bibnamefont {Qiu}}, \bibinfo
  {author} {\bibfnamefont {H.}~\bibnamefont {Song}}, \bibinfo {author}
  {\bibfnamefont {J.}~\bibnamefont {Bernhard}}, \bibinfo {author}
  {\bibfnamefont {S.}~\bibnamefont {Bass}}, \ and\ \bibinfo {author}
  {\bibfnamefont {U.}~\bibnamefont {Heinz}},\ }\href {\doibase
  10.1016/j.cpc.2015.08.039} {\bibfield  {journal} {\bibinfo  {journal}
  {Comput. Phys. Commun.}\ }\textbf {\bibinfo {volume} {199}},\ \bibinfo
  {pages} {61} (\bibinfo {year} {2016}{\natexlab{b}})},\ \Eprint
  {http://arxiv.org/abs/1409.8164} {arXiv:1409.8164 [nucl-th]} \BibitemShut
  {NoStop}%
\bibitem [{\citenamefont {Shen}(7 25)}]{Shen:2014lye}%
  \BibitemOpen
  \bibfield  {author} {\bibinfo {author} {\bibfnamefont {C.}~\bibnamefont
  {Shen}},\ }\emph {\bibinfo {title} {{The standard model for relativistic
  heavy-ion collisions and electromagnetic tomography}}},\ \href
  {http://rave.ohiolink.edu/etdc/view?acc_num=osu1405931790} {Ph.D. thesis},\
  \bibinfo  {school} {Ohio State U.} (\bibinfo {year} {2014-07-25})\BibitemShut
  {NoStop}%
\bibitem [{\citenamefont {Baier}\ \emph {et~al.}(2001)\citenamefont {Baier},
  \citenamefont {Mueller}, \citenamefont {Schiff},\ and\ \citenamefont
  {Son}}]{Baier:2000sb}%
  \BibitemOpen
  \bibfield  {author} {\bibinfo {author} {\bibfnamefont {R.}~\bibnamefont
  {Baier}}, \bibinfo {author} {\bibfnamefont {A.~H.}\ \bibnamefont {Mueller}},
  \bibinfo {author} {\bibfnamefont {D.}~\bibnamefont {Schiff}}, \ and\ \bibinfo
  {author} {\bibfnamefont {D.~T.}\ \bibnamefont {Son}},\ }\href {\doibase
  10.1016/S0370-2693(01)00191-5} {\bibfield  {journal} {\bibinfo  {journal}
  {Phys. Lett.}\ }\textbf {\bibinfo {volume} {B502}},\ \bibinfo {pages} {51}
  (\bibinfo {year} {2001})},\ \Eprint {http://arxiv.org/abs/hep-ph/0009237}
  {arXiv:hep-ph/0009237 [hep-ph]} \BibitemShut {NoStop}%
\bibitem [{\citenamefont {Vogelsang}()}]{VogelsangPrivate}%
  \BibitemOpen
  \bibfield  {author} {\bibinfo {author} {\bibfnamefont {W.}~\bibnamefont
  {Vogelsang}},\ }\href@noop {} {}\bibinfo {howpublished} {Private
  communication}\BibitemShut {NoStop}%
\bibitem [{\citenamefont {Arnold}\ \emph {et~al.}(2001)\citenamefont {Arnold},
  \citenamefont {Moore},\ and\ \citenamefont {Yaffe}}]{Arnold:2001ms}%
  \BibitemOpen
  \bibfield  {author} {\bibinfo {author} {\bibfnamefont {P.~B.}\ \bibnamefont
  {Arnold}}, \bibinfo {author} {\bibfnamefont {G.~D.}\ \bibnamefont {Moore}}, \
  and\ \bibinfo {author} {\bibfnamefont {L.~G.}\ \bibnamefont {Yaffe}},\ }\href
  {\doibase 10.1088/1126-6708/2001/12/009} {\bibfield  {journal} {\bibinfo
  {journal} {JHEP}\ }\textbf {\bibinfo {volume} {12}},\ \bibinfo {pages} {009}
  (\bibinfo {year} {2001})},\ \Eprint {http://arxiv.org/abs/hep-ph/0111107}
  {arXiv:hep-ph/0111107 [hep-ph]} \BibitemShut {NoStop}%
\bibitem [{\citenamefont {Aurenche}\ \emph
  {et~al.}(2000{\natexlab{a}})\citenamefont {Aurenche}, \citenamefont {Gelis},\
  and\ \citenamefont {Zaraket}}]{Aurenche:2000gf}%
  \BibitemOpen
  \bibfield  {author} {\bibinfo {author} {\bibfnamefont {P.}~\bibnamefont
  {Aurenche}}, \bibinfo {author} {\bibfnamefont {F.}~\bibnamefont {Gelis}}, \
  and\ \bibinfo {author} {\bibfnamefont {H.}~\bibnamefont {Zaraket}},\ }\href
  {\doibase 10.1103/PhysRevD.62.096012} {\bibfield  {journal} {\bibinfo
  {journal} {Phys. Rev. D}\ }\textbf {\bibinfo {volume} {62}},\ \bibinfo
  {pages} {096012} (\bibinfo {year} {2000}{\natexlab{a}})}\BibitemShut
  {NoStop}%
\bibitem [{\citenamefont {Aurenche}\ \emph {et~al.}(1998)\citenamefont
  {Aurenche}, \citenamefont {Gelis}, \citenamefont {Zaraket},\ and\
  \citenamefont {Kobes}}]{AURENCHE1}%
  \BibitemOpen
  \bibfield  {author} {\bibinfo {author} {\bibfnamefont {P.}~\bibnamefont
  {Aurenche}}, \bibinfo {author} {\bibfnamefont {F.}~\bibnamefont {Gelis}},
  \bibinfo {author} {\bibfnamefont {H.}~\bibnamefont {Zaraket}}, \ and\
  \bibinfo {author} {\bibfnamefont {R.}~\bibnamefont {Kobes}},\ }\href
  {\doibase 10.1103/PhysRevD.58.085003} {\bibfield  {journal} {\bibinfo
  {journal} {Phys. Rev. D}\ }\textbf {\bibinfo {volume} {58}},\ \bibinfo
  {pages} {085003} (\bibinfo {year} {1998})}\BibitemShut {NoStop}%
\bibitem [{\citenamefont {Aurenche}\ \emph
  {et~al.}(2000{\natexlab{b}})\citenamefont {Aurenche}, \citenamefont {Gelis},\
  and\ \citenamefont {Zaraket}}]{AURENCHE2}%
  \BibitemOpen
  \bibfield  {author} {\bibinfo {author} {\bibfnamefont {P.}~\bibnamefont
  {Aurenche}}, \bibinfo {author} {\bibfnamefont {F.}~\bibnamefont {Gelis}}, \
  and\ \bibinfo {author} {\bibfnamefont {H.}~\bibnamefont {Zaraket}},\ }\href
  {\doibase 10.1103/PhysRevD.61.116001} {\bibfield  {journal} {\bibinfo
  {journal} {Phys. Rev. D}\ }\textbf {\bibinfo {volume} {61}},\ \bibinfo
  {pages} {116001} (\bibinfo {year} {2000}{\natexlab{b}})}\BibitemShut
  {NoStop}%
\bibitem [{\citenamefont {Heffernan}\ \emph {et~al.}(2015)\citenamefont
  {Heffernan}, \citenamefont {Hohler},\ and\ \citenamefont
  {Rapp}}]{Heffernan:2014mla}%
  \BibitemOpen
  \bibfield  {author} {\bibinfo {author} {\bibfnamefont {M.}~\bibnamefont
  {Heffernan}}, \bibinfo {author} {\bibfnamefont {P.}~\bibnamefont {Hohler}}, \
  and\ \bibinfo {author} {\bibfnamefont {R.}~\bibnamefont {Rapp}},\ }\href
  {\doibase 10.1103/PhysRevC.91.027902} {\bibfield  {journal} {\bibinfo
  {journal} {Phys. Rev.}\ }\textbf {\bibinfo {volume} {C91}},\ \bibinfo {pages}
  {027902} (\bibinfo {year} {2015})},\ \Eprint {http://arxiv.org/abs/1411.7012}
  {arXiv:1411.7012 [hep-ph]} \BibitemShut {NoStop}%
\bibitem [{\citenamefont {Rapp}\ and\ \citenamefont
  {Wambach}(1999)}]{Rapp:1999us}%
  \BibitemOpen
  \bibfield  {author} {\bibinfo {author} {\bibfnamefont {R.}~\bibnamefont
  {Rapp}}\ and\ \bibinfo {author} {\bibfnamefont {J.}~\bibnamefont {Wambach}},\
  }\href {\doibase 10.1007/s100500050364} {\bibfield  {journal} {\bibinfo
  {journal} {The European Physical Journal A - Hadrons and Nuclei}\ }\textbf
  {\bibinfo {volume} {6}},\ \bibinfo {pages} {415} (\bibinfo {year}
  {1999})}\BibitemShut {NoStop}%
\bibitem [{\citenamefont {Liu}\ and\ \citenamefont {Rapp}(2007)}]{Liu:2007zzw}%
  \BibitemOpen
  \bibfield  {author} {\bibinfo {author} {\bibfnamefont {W.}~\bibnamefont
  {Liu}}\ and\ \bibinfo {author} {\bibfnamefont {R.}~\bibnamefont {Rapp}},\
  }\href {\doibase https://doi.org/10.1016/j.nuclphysa.2007.08.014} {\bibfield
  {journal} {\bibinfo  {journal} {Nuclear Physics A}\ }\textbf {\bibinfo
  {volume} {796}},\ \bibinfo {pages} {101 } (\bibinfo {year}
  {2007})}\BibitemShut {NoStop}%
\bibitem [{\citenamefont {Khachatryan}\ \emph {et~al.}(2018)\citenamefont
  {Khachatryan}, \citenamefont {Schenke}, \citenamefont {Chiu}, \citenamefont
  {Drees}, \citenamefont {Hemmick},\ and\ \citenamefont
  {Novitzky}}]{Khachatryan:2018ori}%
  \BibitemOpen
  \bibfield  {author} {\bibinfo {author} {\bibfnamefont {V.}~\bibnamefont
  {Khachatryan}}, \bibinfo {author} {\bibfnamefont {B.}~\bibnamefont
  {Schenke}}, \bibinfo {author} {\bibfnamefont {M.}~\bibnamefont {Chiu}},
  \bibinfo {author} {\bibfnamefont {A.}~\bibnamefont {Drees}}, \bibinfo
  {author} {\bibfnamefont {T.~K.}\ \bibnamefont {Hemmick}}, \ and\ \bibinfo
  {author} {\bibfnamefont {N.}~\bibnamefont {Novitzky}},\ }\href {\doibase
  10.1016/j.nuclphysa.2018.07.013} {\bibfield  {journal} {\bibinfo  {journal}
  {Nucl. Phys.}\ }\textbf {\bibinfo {volume} {A978}},\ \bibinfo {pages} {123}
  (\bibinfo {year} {2018})},\ \Eprint {http://arxiv.org/abs/1804.09257}
  {arXiv:1804.09257 [nucl-th]} \BibitemShut {NoStop}%
\bibitem [{\citenamefont {Schenke}\ \emph {et~al.}(2012)\citenamefont
  {Schenke}, \citenamefont {Tribedy},\ and\ \citenamefont
  {Venugopalan}}]{Schenke:2012wb}%
  \BibitemOpen
  \bibfield  {author} {\bibinfo {author} {\bibfnamefont {B.}~\bibnamefont
  {Schenke}}, \bibinfo {author} {\bibfnamefont {P.}~\bibnamefont {Tribedy}}, \
  and\ \bibinfo {author} {\bibfnamefont {R.}~\bibnamefont {Venugopalan}},\
  }\href {\doibase 10.1103/PhysRevLett.108.252301} {\bibfield  {journal}
  {\bibinfo  {journal} {Phys. Rev. Lett.}\ }\textbf {\bibinfo {volume} {108}},\
  \bibinfo {pages} {252301} (\bibinfo {year} {2012})},\ \Eprint
  {http://arxiv.org/abs/1202.6646} {arXiv:1202.6646 [nucl-th]} \BibitemShut
  {NoStop}%
\bibitem [{\citenamefont {Miller}\ \emph {et~al.}(2007)\citenamefont {Miller},
  \citenamefont {Reygers}, \citenamefont {Sanders},\ and\ \citenamefont
  {Steinberg}}]{Miller:2007ri}%
  \BibitemOpen
  \bibfield  {author} {\bibinfo {author} {\bibfnamefont {M.~L.}\ \bibnamefont
  {Miller}}, \bibinfo {author} {\bibfnamefont {K.}~\bibnamefont {Reygers}},
  \bibinfo {author} {\bibfnamefont {S.~J.}\ \bibnamefont {Sanders}}, \ and\
  \bibinfo {author} {\bibfnamefont {P.}~\bibnamefont {Steinberg}},\ }\href
  {\doibase 10.1146/annurev.nucl.57.090506.123020} {\bibfield  {journal}
  {\bibinfo  {journal} {Ann. Rev. Nucl. Part. Sci.}\ }\textbf {\bibinfo
  {volume} {57}},\ \bibinfo {pages} {205} (\bibinfo {year} {2007})},\ \Eprint
  {http://arxiv.org/abs/nucl-ex/0701025} {arXiv:nucl-ex/0701025 [nucl-ex]}
  \BibitemShut {NoStop}%
\bibitem [{\citenamefont {Kowalski}\ and\ \citenamefont
  {Teaney}(2003)}]{Kowalski:2003hm}%
  \BibitemOpen
  \bibfield  {author} {\bibinfo {author} {\bibfnamefont {H.}~\bibnamefont
  {Kowalski}}\ and\ \bibinfo {author} {\bibfnamefont {D.}~\bibnamefont
  {Teaney}},\ }\href {\doibase 10.1103/PhysRevD.68.114005} {\bibfield
  {journal} {\bibinfo  {journal} {Phys. Rev.}\ }\textbf {\bibinfo {volume}
  {D68}},\ \bibinfo {pages} {114005} (\bibinfo {year} {2003})},\ \Eprint
  {http://arxiv.org/abs/hep-ph/0304189} {arXiv:hep-ph/0304189 [hep-ph]}
  \BibitemShut {NoStop}%
\bibitem [{\citenamefont {Rezaeian}\ \emph {et~al.}(2013)\citenamefont
  {Rezaeian}, \citenamefont {Siddikov}, \citenamefont {Van~de Klundert},\ and\
  \citenamefont {Venugopalan}}]{Rezaeian:2012ji}%
  \BibitemOpen
  \bibfield  {author} {\bibinfo {author} {\bibfnamefont {A.~H.}\ \bibnamefont
  {Rezaeian}}, \bibinfo {author} {\bibfnamefont {M.}~\bibnamefont {Siddikov}},
  \bibinfo {author} {\bibfnamefont {M.}~\bibnamefont {Van~de Klundert}}, \ and\
  \bibinfo {author} {\bibfnamefont {R.}~\bibnamefont {Venugopalan}},\ }\href
  {\doibase 10.1103/PhysRevD.87.034002} {\bibfield  {journal} {\bibinfo
  {journal} {Phys. Rev.}\ }\textbf {\bibinfo {volume} {D87}},\ \bibinfo {pages}
  {034002} (\bibinfo {year} {2013})},\ \Eprint {http://arxiv.org/abs/1212.2974}
  {arXiv:1212.2974 [hep-ph]} \BibitemShut {NoStop}%
\bibitem [{\citenamefont {Garcia-Montero}(2019)}]{Garcia-Montero:2019vju}%
  \BibitemOpen
  \bibfield  {author} {\bibinfo {author} {\bibfnamefont {O.}~\bibnamefont
  {Garcia-Montero}},\ }\href@noop {} {\  (\bibinfo {year} {2019})},\ \Eprint
  {http://arxiv.org/abs/1909.12294} {arXiv:1909.12294 [hep-ph]} \BibitemShut
  {NoStop}%
\bibitem [{\citenamefont {Berges}\ \emph
  {et~al.}(2014{\natexlab{a}})\citenamefont {Berges}, \citenamefont
  {Boguslavski}, \citenamefont {Schlichting},\ and\ \citenamefont
  {Venugopalan}}]{Berges:2013eia}%
  \BibitemOpen
  \bibfield  {author} {\bibinfo {author} {\bibfnamefont {J.}~\bibnamefont
  {Berges}}, \bibinfo {author} {\bibfnamefont {K.}~\bibnamefont {Boguslavski}},
  \bibinfo {author} {\bibfnamefont {S.}~\bibnamefont {Schlichting}}, \ and\
  \bibinfo {author} {\bibfnamefont {R.}~\bibnamefont {Venugopalan}},\ }\href
  {\doibase 10.1103/PhysRevD.89.074011} {\bibfield  {journal} {\bibinfo
  {journal} {Phys. Rev.}\ }\textbf {\bibinfo {volume} {D89}},\ \bibinfo {pages}
  {074011} (\bibinfo {year} {2014}{\natexlab{a}})},\ \Eprint
  {http://arxiv.org/abs/1303.5650} {arXiv:1303.5650 [hep-ph]} \BibitemShut
  {NoStop}%
\bibitem [{\citenamefont {Berges}\ \emph
  {et~al.}(2014{\natexlab{b}})\citenamefont {Berges}, \citenamefont
  {Boguslavski}, \citenamefont {Schlichting},\ and\ \citenamefont
  {Venugopalan}}]{Berges:2013lsa}%
  \BibitemOpen
  \bibfield  {author} {\bibinfo {author} {\bibfnamefont {J.}~\bibnamefont
  {Berges}}, \bibinfo {author} {\bibfnamefont {K.}~\bibnamefont {Boguslavski}},
  \bibinfo {author} {\bibfnamefont {S.}~\bibnamefont {Schlichting}}, \ and\
  \bibinfo {author} {\bibfnamefont {R.}~\bibnamefont {Venugopalan}},\ }\href
  {\doibase 10.1007/JHEP05(2014)054} {\bibfield  {journal} {\bibinfo  {journal}
  {JHEP}\ }\textbf {\bibinfo {volume} {05}},\ \bibinfo {pages} {054} (\bibinfo
  {year} {2014}{\natexlab{b}})},\ \Eprint {http://arxiv.org/abs/1312.5216}
  {arXiv:1312.5216 [hep-ph]} \BibitemShut {NoStop}%
\bibitem [{\citenamefont {Berges}\ \emph
  {et~al.}(2014{\natexlab{c}})\citenamefont {Berges}, \citenamefont
  {Boguslavski}, \citenamefont {Schlichting},\ and\ \citenamefont
  {Venugopalan}}]{Berges:2013fga}%
  \BibitemOpen
  \bibfield  {author} {\bibinfo {author} {\bibfnamefont {J.}~\bibnamefont
  {Berges}}, \bibinfo {author} {\bibfnamefont {K.}~\bibnamefont {Boguslavski}},
  \bibinfo {author} {\bibfnamefont {S.}~\bibnamefont {Schlichting}}, \ and\
  \bibinfo {author} {\bibfnamefont {R.}~\bibnamefont {Venugopalan}},\ }\href
  {\doibase 10.1103/PhysRevD.89.114007} {\bibfield  {journal} {\bibinfo
  {journal} {Phys. Rev.}\ }\textbf {\bibinfo {volume} {D89}},\ \bibinfo {pages}
  {114007} (\bibinfo {year} {2014}{\natexlab{c}})},\ \Eprint
  {http://arxiv.org/abs/1311.3005} {arXiv:1311.3005 [hep-ph]} \BibitemShut
  {NoStop}%
\bibitem [{\citenamefont {Blaizot}\ \emph {et~al.}(2014)\citenamefont
  {Blaizot}, \citenamefont {Wu},\ and\ \citenamefont {Yan}}]{Blaizot:2014jna}%
  \BibitemOpen
  \bibfield  {author} {\bibinfo {author} {\bibfnamefont {J.-P.}\ \bibnamefont
  {Blaizot}}, \bibinfo {author} {\bibfnamefont {B.}~\bibnamefont {Wu}}, \ and\
  \bibinfo {author} {\bibfnamefont {L.}~\bibnamefont {Yan}},\ }\href {\doibase
  10.1016/j.nuclphysa.2014.07.041} {\bibfield  {journal} {\bibinfo  {journal}
  {Nucl. Phys.}\ }\textbf {\bibinfo {volume} {A930}},\ \bibinfo {pages} {139}
  (\bibinfo {year} {2014})},\ \Eprint {http://arxiv.org/abs/1402.5049}
  {arXiv:1402.5049 [hep-ph]} \BibitemShut {NoStop}%
\bibitem [{\citenamefont {Lappi}(2008)}]{Lappi:2007ku}%
  \BibitemOpen
  \bibfield  {author} {\bibinfo {author} {\bibfnamefont {T.}~\bibnamefont
  {Lappi}},\ }\href {\doibase 10.1140/epjc/s10052-008-0588-4} {\bibfield
  {journal} {\bibinfo  {journal} {Eur. Phys. J.}\ }\textbf {\bibinfo {volume}
  {C55}},\ \bibinfo {pages} {285} (\bibinfo {year} {2008})},\ \Eprint
  {http://arxiv.org/abs/0711.3039} {arXiv:0711.3039 [hep-ph]} \BibitemShut
  {NoStop}%
\bibitem [{\citenamefont {Kapusta}\ \emph {et~al.}(1991)\citenamefont
  {Kapusta}, \citenamefont {Lichard},\ and\ \citenamefont {Seibert}}]{KAPUSTA}%
  \BibitemOpen
  \bibfield  {author} {\bibinfo {author} {\bibfnamefont {J.}~\bibnamefont
  {Kapusta}}, \bibinfo {author} {\bibfnamefont {P.}~\bibnamefont {Lichard}}, \
  and\ \bibinfo {author} {\bibfnamefont {D.}~\bibnamefont {Seibert}},\ }\href
  {\doibase 10.1103/PhysRevD.44.2774} {\bibfield  {journal} {\bibinfo
  {journal} {Phys. Rev. D}\ }\textbf {\bibinfo {volume} {44}},\ \bibinfo
  {pages} {2774} (\bibinfo {year} {1991})},\ \bibinfo {note} {erratum-ibid.
  ~\cite{KAPUSTAerratum}}\BibitemShut {NoStop}%
\bibitem [{\citenamefont {Paquet}(2017)}]{Paquet:2017wji}%
  \BibitemOpen
  \bibfield  {author} {\bibinfo {author} {\bibfnamefont {J.-F.}\ \bibnamefont
  {Paquet}},\ }\bibfield  {booktitle} {\emph {\bibinfo {booktitle}
  {{Proceedings, 26th International Conference on Ultra-relativistic
  Nucleus-Nucleus Collisions (Quark Matter 2017): Chicago, Illinois, USA,
  February 5-11, 2017}}},\ }\href {\doibase 10.1016/j.nuclphysa.2017.06.003}
  {\bibfield  {journal} {\bibinfo  {journal} {Nucl. Phys.}\ }\textbf {\bibinfo
  {volume} {A967}},\ \bibinfo {pages} {184} (\bibinfo {year} {2017})},\ \Eprint
  {http://arxiv.org/abs/1704.07842} {arXiv:1704.07842 [nucl-th]} \BibitemShut
  {NoStop}%
\bibitem [{\citenamefont {Rapp}(2013)}]{Rapp:2013ema}%
  \BibitemOpen
  \bibfield  {author} {\bibinfo {author} {\bibfnamefont {R.}~\bibnamefont
  {Rapp}},\ }\bibfield  {booktitle} {\emph {\bibinfo {booktitle} {{Proceedings,
  8th International Workshop on Critical Point and Onset of Deconfinement (CPOD
  2013): Napa, CA, USA, March 11-15, 2013}}},\ }\href {\doibase
  10.22323/1.185.0008} {\bibfield  {journal} {\bibinfo  {journal} {PoS}\
  }\textbf {\bibinfo {volume} {CPOD2013}},\ \bibinfo {pages} {008} (\bibinfo
  {year} {2013})},\ \Eprint {http://arxiv.org/abs/1306.6394} {arXiv:1306.6394
  [nucl-th]} \BibitemShut {NoStop}%
\bibitem [{\citenamefont {Shen}\ \emph {et~al.}(2014)\citenamefont {Shen},
  \citenamefont {Heinz}, \citenamefont {Paquet},\ and\ \citenamefont
  {Gale}}]{Shen:2013vja}%
  \BibitemOpen
  \bibfield  {author} {\bibinfo {author} {\bibfnamefont {C.}~\bibnamefont
  {Shen}}, \bibinfo {author} {\bibfnamefont {U.~W.}\ \bibnamefont {Heinz}},
  \bibinfo {author} {\bibfnamefont {J.-F.}\ \bibnamefont {Paquet}}, \ and\
  \bibinfo {author} {\bibfnamefont {C.}~\bibnamefont {Gale}},\ }\href {\doibase
  10.1103/PhysRevC.89.044910} {\bibfield  {journal} {\bibinfo  {journal} {Phys.
  Rev.}\ }\textbf {\bibinfo {volume} {C89}},\ \bibinfo {pages} {044910}
  (\bibinfo {year} {2014})},\ \Eprint {http://arxiv.org/abs/1308.2440}
  {arXiv:1308.2440 [nucl-th]} \BibitemShut {NoStop}%
\bibitem [{\citenamefont {Kaczmarek}\ and\ \citenamefont
  {Zantow}(2005)}]{Kaczmarek:2005ui}%
  \BibitemOpen
  \bibfield  {author} {\bibinfo {author} {\bibfnamefont {O.}~\bibnamefont
  {Kaczmarek}}\ and\ \bibinfo {author} {\bibfnamefont {F.}~\bibnamefont
  {Zantow}},\ }\href {\doibase 10.1103/PhysRevD.71.114510} {\bibfield
  {journal} {\bibinfo  {journal} {Phys. Rev.}\ }\textbf {\bibinfo {volume}
  {D71}},\ \bibinfo {pages} {114510} (\bibinfo {year} {2005})},\ \Eprint
  {http://arxiv.org/abs/hep-lat/0503017} {arXiv:hep-lat/0503017 [hep-lat]}
  \BibitemShut {NoStop}%
\bibitem [{\citenamefont {Weldon}(1982)}]{Weldon:1982bn}%
  \BibitemOpen
  \bibfield  {author} {\bibinfo {author} {\bibfnamefont {H.~A.}\ \bibnamefont
  {Weldon}},\ }\href {\doibase 10.1103/PhysRevD.26.2789} {\bibfield  {journal}
  {\bibinfo  {journal} {Phys. Rev.}\ }\textbf {\bibinfo {volume} {D26}},\
  \bibinfo {pages} {2789} (\bibinfo {year} {1982})}\BibitemShut {NoStop}%
\bibitem [{\citenamefont {Kapusta}\ \emph {et~al.}(1993)\citenamefont
  {Kapusta}, \citenamefont {Lichard},\ and\ \citenamefont
  {Seibert}}]{KAPUSTAerratum}%
  \BibitemOpen
  \bibfield  {author} {\bibinfo {author} {\bibfnamefont {J.}~\bibnamefont
  {Kapusta}}, \bibinfo {author} {\bibfnamefont {P.}~\bibnamefont {Lichard}}, \
  and\ \bibinfo {author} {\bibfnamefont {D.}~\bibnamefont {Seibert}},\ }\href
  {\doibase 10.1103/PhysRevD.47.4171} {\bibfield  {journal} {\bibinfo
  {journal} {Phys. Rev. D}\ }\textbf {\bibinfo {volume} {47}},\ \bibinfo
  {pages} {4171} (\bibinfo {year} {1993})}\BibitemShut {NoStop}%
\end{thebibliography}%

\end{document}